\documentclass[twocolumn]{aastex61}
\usepackage{graphicx,natbib}
\usepackage{upgreek}
\usepackage[figuresright]{rotating}

\usepackage{soul} 
\usepackage{amsmath}

\newcommand\HI{H{\sc i}}
\newcommand\spiral{spiral sample}
\newcommand\Mss{M$_s$ sample}
\newcommand\Ms{$M_\ast$}
\newcommand\Msolar{${\rm M_\sun}$}
\newcommand\Mh{$M_{{H{\sc I}}}$}
\newcommand\tdep{t$_{\mathrm{dep}}$}

\shorttitle{HI-WISE}
\shortauthors{Parkash et al.}

\date{\today}

\begin{document}

\title{Relationships between HI Gas Mass, Stellar Mass and Star Formation Rate of HICAT+WISE (HI-WISE) Galaxies}

\correspondingauthor{Vaishali Parkash}
\email{vaishali.parkash@monash.edu}

\author{Vaishali Parkash}
\affil{ School of Physics and Astronomy, Monash Centre for Astrophysics (MoCA), Monash University, Clayton, Victoria 3800, Australia}

\author{Michael J. I. Brown}
\affil{ School of Physics and Astronomy, Monash Centre for Astrophysics (MoCA), Monash University, Clayton, Victoria 3800, Australia}

\author{T. H. Jarrett}
\affil{Astrophysics, Cosmology and Gravity Centre (ACGC), Astronomy Department, University of Cape Town, Private Bag X3, Rondebosch 7701, South Africa}

\author{Nicolas J. Bonne}
\affil{Institute for Cosmology and Gravitation, Dennis Sciama Building, University of Portsmouth, Burnaby Road, Portsmouth PO1 3FX, United Kingdom}

\begin{abstract}
We have measured the relationships between \HI\ mass, stellar mass and star formation rate using the \HI\ Parkes All Sky-Survey Catalogue (HICAT) and the Wide-field Infrared Survey Explorer (WISE). Of the 3,513 HICAT sources, we find 3.4 $\upmu$m counterparts for 2,896 sources (80\%) and provide new WISE matched aperture photometry for these galaxies. For our principal sample of spiral galaxies with $W1$ $\le$ 10 mag and $z$ $\le$ 0.01, we identify \HI\ detections for 93\% of the sample. We measure lower \HI\--stellar mass relationships that \HI\ selected samples that do not include spiral galaxies with little \HI\ gas. Our observations of the \spiral\ show that \HI\ mass increases with stellar mass with a power-law index 0.35; however, this value is dependent on T-type, which affects both the median and the dispersion of \HI\ mass. We also observe an upper limit on the \HI\ gas fraction, which is consistent with a halo spin parameter model. We measure the star formation efficiency of spiral galaxies to be constant at $10^{-9.57}$ yr$^{-1}$ $\pm$ 0.4 dex for 2.5 orders of magnitude in stellar mass, despite the higher stellar mass spiral showing evidence of quenched star formation. 
 \end{abstract}
 \keywords{galaxies: evolution -- galaxies: fundamental parameters -- galaxies: spiral -- galaxies: star formation  --  radio lines: galaxies -- radio lines: ISM }

\section{Introduction} \label{sec:intro}

Star formation is fueled by atomic (\HI) and molecular hydrogen, so we expect correlations between \HI\ mass, stellar mass and star formation rates (SFR). This is exemplified by the Kennicutt--Schmidt law that establishes a correlation between gas surface density and SFR, albeit both for \HI\ and molecular gas, and with considerable scatter \citep{kennicutt98}. Correlations are also observed between stellar mass and \HI\ mass \citep[e.g.,][]{catinella10, huang12, maddox15}, although these are partially explained by luminosity--luminosity correlations. However, the presence of mass quenching indicates that \HI\ mass declines above some high stellar mass threshold \citep[e.g.,][]{kauff03,gabor10}.

Although the relationship between \HI\ mass and SFR in a galaxy is predicted by the Kennicutt--Schmidt law, there is a scatter of $\sim$0.2 dex \citep{kennicutt98}, implying that \HI\ mass is just one of several factors that regulate SFR. Multiple studies using different surveys find that \HI\ mass (\Mh) increases with SFR \citep{mirabel88,doyle06}. \HI\ disk is a precondition for star formation, but star formation also requires gas accretion from the intergalactic medium to drive the \HI\ gas inward, thereby cooling and converting it into molecular hydrogen \citep[e.g.,][]{prochaska09, obreschkow16}. By comparison, the relationship between molecular hydrogen and SFR is better understood since stars form in molecular clouds. Recent works \citep{leroy08, schruba11} have shown that molecular hydrogen, rather than \HI, drives the Kennicutt--Schmidt law, but an empirical study on the relationship between molecular hydrogen and SFR using a large sample of galaxies is difficult to perform because of the lack of large-scale CO surveys. Also, there is uncertainty in converting CO intensity to molecular hydrogen content due to the uncertainty in the X-factor \citep{bolatto13}. Although \HI\ gas is only indirectly related to the SFR, the measurement of \HI\ gas can imply information about the molecular hydrogen of a galaxy \citep[e.g.,][]{leroy08,wong16}.

Multiple studies have found that \HI\ mass increases with stellar mass \citep{catinella10, huang12, maddox15}, which is partially explained by luminosity--luminosity correlations. However, the relations measured by different studies disagree with each other. \citet{huang12} find $\langle\log{}$ \Mh$\rangle$ $\propto0.712\langle\log{}$ \Ms $\rangle$ for \Ms $\le$ 10$^9$ \Msolar\ and $\langle\log{}$ \Mh$\rangle$ $\propto0.276\langle\log{}$ \Ms $\rangle$ for \Ms\ $>$ 10$^9$ \Msolar. The break in the relationship represents the transition from irregular, low stellar mass galaxies to high mass, disk galaxies \citep{maddox15, keres09}. However, \citet{catinella10} observed that \HI\ mass increases little with stellar mass (to the power of 0.02) for galaxies with \Ms\ $>$ 10$^{9}$ \Msolar.

The differences in the relation between \HI\ mass and stellar mass in the literature may be due to sample selection. The samples of \citet{huang12} and \citet{maddox15} use the \HI\ Arecibo Legacy Fast ALFA survey (ALFALFA) population, while the GASS sample of \citet{catinella10} is selected by stellar mass (\Ms\ $> 10^{10}$ \Msolar). By definition, the \HI\ selected samples of \citet{huang12} and \citet{maddox15} are more \HI\--rich compared to a stellar mass-selected sample, and are biased against high stellar mass elliptical galaxies with very little \HI\ gas, resulting in elevated trends for \HI\ mass versus stellar masses. While the literature shows \HI\ mass increases with stellar mass, perhaps with a break or a plateau at high stellar masses \citep[i.e.,][]{catinella10, huang12, maddox15}, there are quantitative disagreements that we suspect are the result of sample selections.

How the \HI\ based star formation efficiency (SFE), or its inverse, the \HI\ depletion timescale (\tdep), varies with a galaxy's stellar mass is unclear from the current literature. \citet{huang12}, \citet{jaskot15}, and \citet{lutz17} find that SFE increases with stellar mass for \HI\ selected samples. Therefore, low mass galaxies that are more \HI-rich than high stellar mass galaxies are inefficient at using their fuel reservoirs to form stars. Contradicting this, \citet{schiminovich10} find SFE of massive galaxies (\Ms $>$ 10$^{10}$ \Msolar) to be constant at 10$^{-9.5}$ yr$^{-1}$, and \citet{wong16} find SFE to be constant at 10$^{-9.65}$ yr$^{-1}$ across 5 orders of magnitude of stellar mass for star-forming galaxies. Incompleteness and sample size are issues for relations from the prior literature, with \citet{jaskot15} having WISE detections for 63\% of their \HI\ sources while \citet{wong16} is highly complete, but contains just 84 galaxies. 

A key limitation of previous studies is low completeness, particularly for infrared counterparts for \HI\ sources, which facilitate the measurement of stellar masses and SFRs. \citet{doyle06} used fluxes from Infrared Astronomical Satellite (IRAS) to calculate SFRs for galaxies in the \HI\ Parkes All-Sky Survey (HIPASS) optical catalog, HOPCAT \citet{doyle05}. Due to the $0.5^{\prime}$ angular resolution and 0.7 Jy 10-sigma sensitivity of IRAS at 12 ${\rm\upmu m}$, they only found infrared counterparts for 32\% of the \HI\ Parkes All-Sky Survey catalog (HICAT). Their final sample comprised of galaxies with high SFRs and excluded galaxies with low rates of star formation, including elliptical galaxies and dwarf galaxies, because at $z$ $=$ 0.01, IRAS can only detect sources brighter than $\nu L_{\nu}$ $\sim1\times10^{10}$ ${\rm L_{\odot}}$. \citet{jaskot15} improved on previous studies by using WISE, which has a 12 ${\rm \upmu m}$ 5-sigma sensitivity of 1 mJy, but at the $z\sim0.06$ maximum redshift of ALFALFA galaxies WISE detects galaxies brighter than $\nu L_{\nu}$ $\sim$ 6$\times10^{8}$ ${\rm L_{\odot}}$, which resulted in \cite{jaskot15} finding infrared counterparts for just 63\% of their \HI\ sources. To improve on the prior literature, we need large samples of galaxies that are highly complete for \HI\ counterparts, while probing large ranges of stellar mass, \HI\ mass, and SFR. 
 
For this work, we measure the relationship between \HI\ mass, stellar mass, and SFR using three galaxy samples combining HICAT and WISE. The paper is arranged as follows: Section \ref{data} describes data used in this research, Section \ref{measurable} details the equations used to calculate the masses and SFRs, Section \ref{samples} describes the samples, Sections \ref{MassRelations} and \ref{SFR} present the results, Section \ref{Discussion} discuss the results and Section \ref{summary} summarizes our work. All magnitudes are in the Vega system. The cosmology applied in this paper is H$_0$ $=$ 70 km s$^{-1}$, $\Omega_M$ $=$ 0.3, and $\Omega_\Lambda$ $=$ 0.7.

\section{Data}\label{data}

\subsection{HICAT}

The principal source of data for our analysis is the HICAT catalog, which is derived from the HIPASS survey \citep{barnes01,meyer04}. HIPASS is a blind survey below a declination $\delta$ of +2$\degr$, performed with the Parkes 64-m radio telescope using a 21 cm multi-beam receiver. The HICAT catalog contains 4,315 \HI\ sources selected from HIPASS and is 99\% complete at a peak flux of 84 mJy and an integrated flux of 9.4 Jy km s$^{-1}$ \citep{meyer04, zwaan04}. 

For each \HI\ source in HICAT, we search for optical counterparts using the position and velocity measurements in the HICAT catalog. Since the HIPASS data has a spatial resolution of 15.5\arcmin, it is necessary to obtain more accurate positions for the \HI\ sources before searching for their mid-infrared counterparts in the WISE frames. To this end, we search for optical counterparts in the spectroscopic sample of \citet{bonne15}, the NASA/IPAC Extragalactic Database (NED)\footnote{The NASA/IPAC Extragalactic Database (NED) is operated by the Jet Propulsion Laboratory, California Institute of Technology, under contract with the National Aeronautics and Space Administration.}, HYPERLEDA \citep{makarov14}, and HOPCAT \citep{doyle05}. In order to accurately match the \HI\ sources with their optical counterparts, we cross-check the velocity measurements from HICAT with the velocity provided from the above sources for all possible positional matches. We describe each source below---as well as the matching process---in detail.
 
In our preliminary search for optical counterparts, we use the spectroscopic sample of \citet{bonne15}, which covers the full survey area of HICAT and provides velocities \citep[predominantly from the r$_{F}$ $\le$ 15.60 2MASS selected 6dF Galaxy Survey, 6dFGS;][]{jones09}, morphologies, and WISE photometry from the All-Sky public-release archive \citep{cutri12} for 13,325 galaxies. For each source, we select the best match to this sample within 5\arcmin\ and 400 km s$^{-1}$ of the respective HICAT position and velocity, resulting in 1,043 optical counterparts of 4,315 \HI\ sources.

To obtain additional optical counterparts, we then search in NED and HYPERLEDA for galaxies within 10\arcmin\ and 400 km s$^{-1}$ of their respective HICAT positions and velocities. As NED and HYPERLEDA draw information from catalogs such as HICAT, it must be ensured that the velocities extracted from these databases are sourced from optical or high-resolution \HI\ radio observations and not sourced from HICAT, and hence that HICAT velocities are not being cross-matched to themselves.

HOPCAT \citep{doyle05} is used as our last source for optical matches. The matches from HOPCAT can not be cross-checked with known velocities, but are categorized as `good guesses' by \citet{doyle05} (see the reference for details on HOPCAT and its matching criteria). We do not use HOPCAT as our primary source for optical matches because \citet{bonne15}, NED, and HYPERLEDA described above provide more recent velocity measurements from 6dFGS and other surveys that were not available at the time HOPCAT was compiled. The final number of galaxies taken from each source described here is listed in Table \ref{table:final}. In total, optical counterparts were obtained for 3,719 \HI\ sources. However, 147 of these were found to be multi-galaxy systems and were thus excluded from our analysis. 

\begin{table}[htb!]
\centering
\caption{The number of optical matches for HICAT.}
\label{table:final}
\scalebox{0.9}{
\begin{tabular}{ll}
\hline\hline
& Number of sources \\ 
 \hline 
Total & 4,315 \\ \hline
No optical match & 596 \\
Position and velocity match & 3,313\\  
Position match & 406\\ \hline
\citet{bonne15} & 1043\\
NED and HYPERLEDA & 2,270\\ 
HOPCAT & 406\\ \hline
\end{tabular}
}
\end{table}

\subsection{WISE Photometry} \label{photometry}

WISE was launched in December 2009, and mapped the entire sky in four mid-infrared bands: W1, W2, W3 and W4 \citep[3.4 $\upmu$m, 4.6 $\upmu$m, 12 $\upmu$m, and 22 $\upmu$m;][]{wright10}. Once WISE depleted its cryogen in October 2010, it was then operated in a ``warm" state using the two short bands and then placed in hibernation for well over 2 years. As part of the NEOWISE program, WISE was reactivated in September 2013 and continues to observe in the W1 and W2 bands \citep{mainzer14}. The point source sensitivities of W1, W2, W3 and W4 in Vega magnitudes are 16.5, 15.5, 11.2 and 7.9, respectively \citep{wright10}, in which W4 is approximately two orders of magnitude more sensitive than IRAS.

We have measured new WISE photometry for the optical counterparts of the HICAT source using the procedure described in \citet{jarrett13}. We chose to do this because the profile-fit photometry data in the ALLWISE public-release archive \citep{cutri12} of the (degraded resolution) mosaics is optimized for point sources \citep[e.g.,][and references therein]{jarrett13, cluver14}. Therefore, resolved sources such as our sample galaxies are either measured as several point sources, or their flux is underestimated by the PSF photometry (mpro). The WISE default catalogs do include extended source photometry for 2MASS Extended Source Catalog \citep[2MXSC;][]{jarrett00} galaxies, but the elliptical apertures (gmag) misses a significant fraction of the flux \citep{cluver14}.

All measurements are carried out on WISE image mosaics that are constructed from single native frames using a drizzle technique \citep{jarrett12}, re-sampled with 1 sq. arc pixels (relative to a 6 arcsec FWHM beam). Photometry for each individual HIPASS galaxy, principally flux measurements and surface brightness characterizations, are conducted using the system developed by T.H Jarrett specifically for WISE data \citep[][see Section 3.6]{jarrett13}. The system estimates photometric errors from the formal components, including the sky background variance and the local sky level, instrumental signatures and the absolute calibration. The error model also takes into account the correlation between re-sampled pixels through a correction factor, which is detailed in the WISE Explanatory Supplement \citep[see Section 2.3.f,][]{cutri12}. As detailed in \citet{jarrett13}, the shape (inclination) and orientation were determined at a fixed $3\sigma$ isophotal level, which provides a robust and relatively accurate ($<$ 5\%) estimate, although this assumes symmetry and a fixed shape to the 1-sigma edge of the galaxy.
 
As detailed in \citet[][2018, in prep]{jarrett13}, isophotal measurements at the W1 1-sigma level typically capture more than 96\% of the total light for bulge-dominated galaxies, and to a lesser extent ($\ge$ 90\%) for late-type galaxies, and most notably in the W3 and W4 bands as much as 20\% of the light can be missing with low surface brightness galaxies. Hence, total fluxes are important in order to estimate the dust-obscured star formation activity in the W3 and W4 bands. The total flux is estimated by fitting a double Sersic function to the axisymmetric radial profile, consisting of an inner bulge and an outer disk. Integrating the composite Sersic model from the $1\sigma$ isophote to the edge of the galaxy (3 disk scale lengths) recovers the light that is below the single-pixel noise threshold; details of the fitting process in \citet{jarrett13}. The error model for the total fluxes includes the goodness of fit, as well as the previous sky estimation per pixel estimates, and typically adds 4 to 5\% to the isophotal flux uncertainty.

Lastly, each galaxy mosaic is visually inspected, and if necessary, bright stars and nearby galaxies are manually masked out, and the apertures are adjusted. Figure 1 compares our new WISE photometry for HICAT galaxies with the previously available archival fluxes and illustrates the impact of the new photometry on derived quantities. For example, for W1$\sim$12 mag galaxies our W1 and W3 magnitudes are on average systematically brighter by $\sim$ 1.4 mag and 1.1 mag than the default photometry pipeline magnitudes. Figure \ref{fig:snapshots} shows four examples of the WISE mosaics that have been cleaned of neighboring objects, as well as the elliptical apertures used for photometry.

\begin{figure}[htb!]
\center
\includegraphics[width= 3in,keepaspectratio]{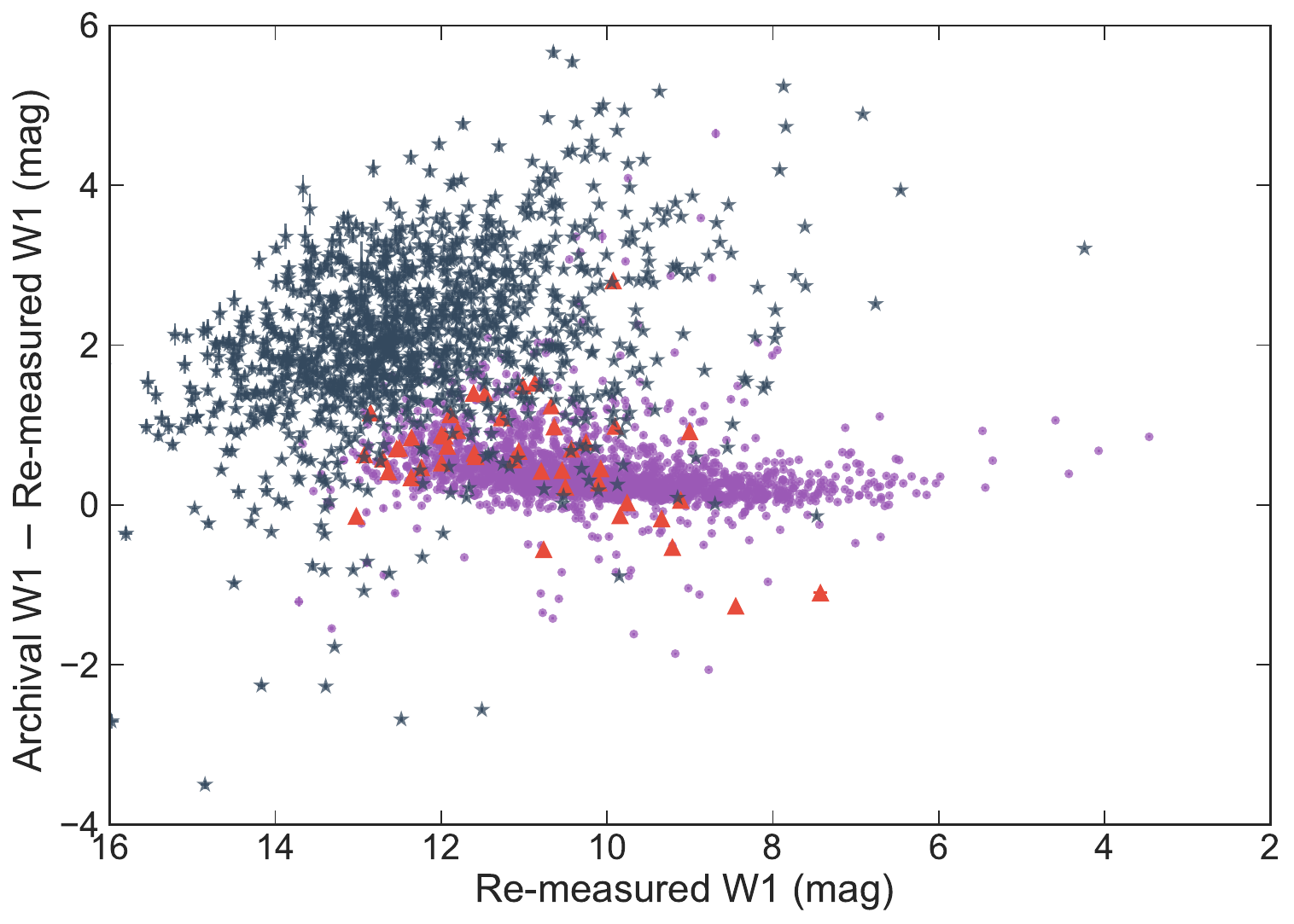}
\includegraphics[width= 3in,keepaspectratio]{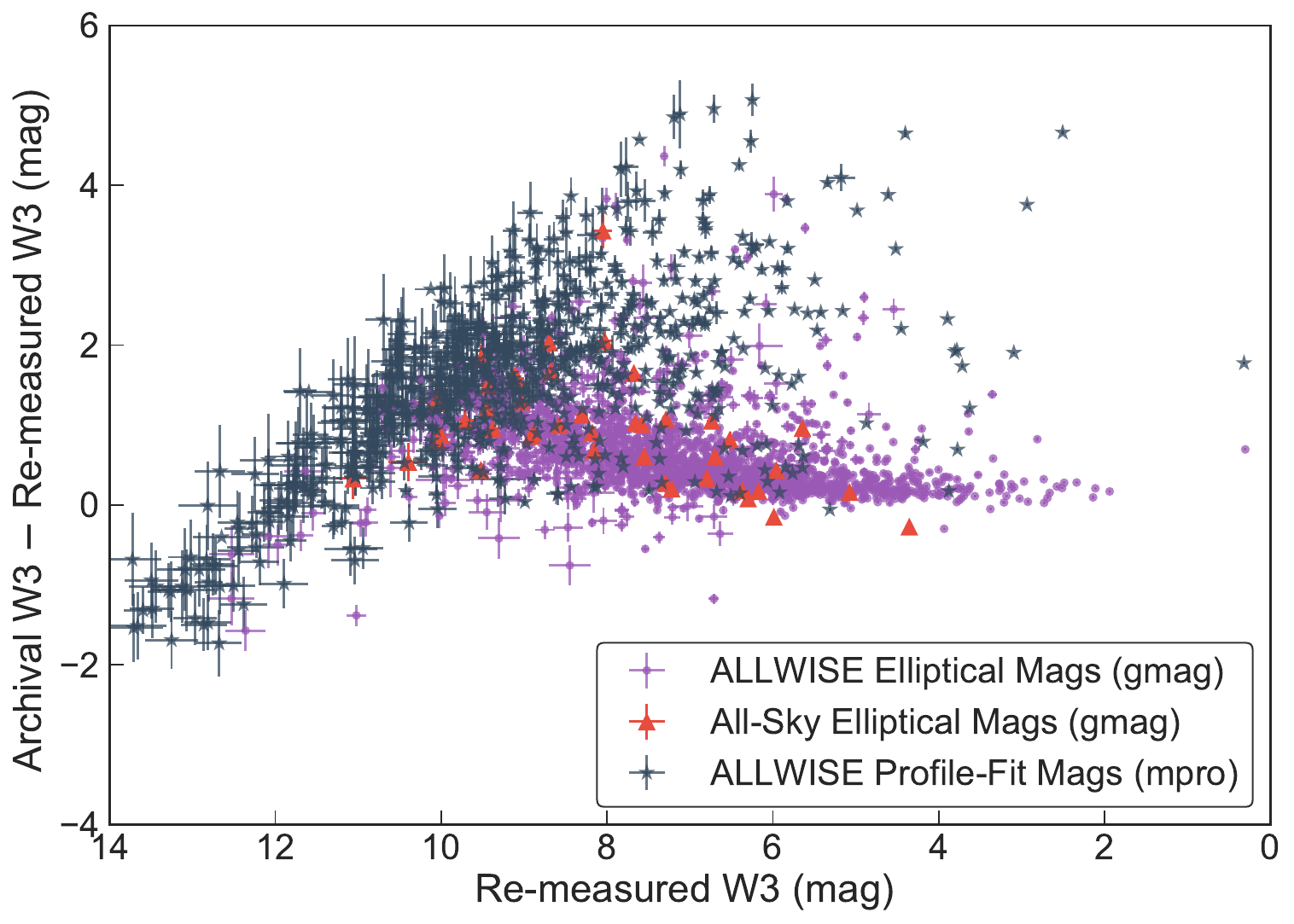}
\caption{The difference between the archival and new magnitudes as a function of the new magnitude for W1 (top) and W3 (bottom). In order of preference, the archival data are either elliptical aperture magnitudes from the ALLWISE (purple) catalog or All-Sky magnitudes (navy) catalog, or PSF profile-fit (red) magnitudes from the ALLWISE catalog \citep{cutri13}. For galaxies with W1$\sim$12 mag, the default elliptical aperture photometry is typically in error by $\sim$1 magnitude, while the PSF profile-fit photometry is in error by $\sim$2 magnitudes. }
\label{fig:phot}
\end{figure}

\begin{figure}[htb!]
\plotone{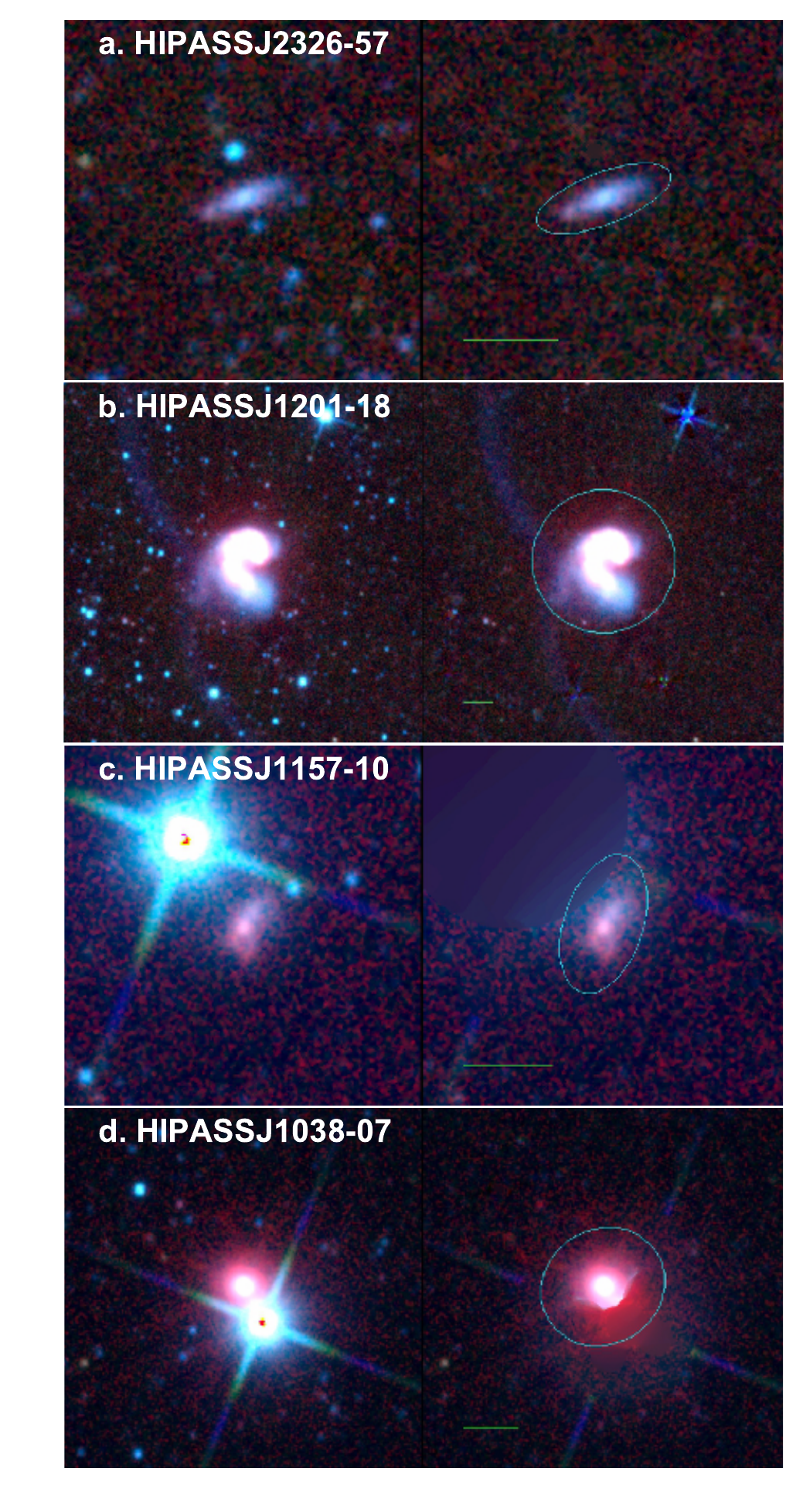}
\caption{WISE color images image composed of W1, W2, W3, W4 bands of four \HI\ sources showing the galaxies before (left) and after (right) star and background source removal. The cyan ellipses indicate the $1\sigma$ apertures. The light from evolved stars are in blue and active star formation is in red. A scale of 2\arcmin\ is indicated with the green horizontal line. North is upwards and East is to the left. HIPASSJ2326-57 is an example of a well behaved photometric galaxy; HIPASSJ1201-18 is a multi-galaxy \HI\ source; HIPASSJ1157-10 and HIPASSJ1038-07 are both flagged visually and have a W1 and W2 S/N $<$ 5 respectively.}
\label{fig:snapshots}
\end{figure}

We apply a S/N threshold of 5 in the W1 and W2 bands, a S/N threshold of 3 in the W3 band, and reject confused sources or \HI\ sources consisting of multiple galaxies. Consequently, we measure good W1-W2 photometry for 3,275 \HI\ sources and good W1-W2-W3 photometry for 2,831 \HI\ sources. We find 20 \HI\ sources that do not meet any of our WISE signal-to-noise thresholds and 147 \HI\ sources are multi-galaxy sources. From top to bottom, Figure \ref{fig:snapshots} shows examples from HICAT of a ``well behaved" source, a multi-galaxy source, and two visually flagged sources. A full list of parameters for HICAT+WISE (HI-WISE) is given in Table \ref{tab:param_desc} in the Appendix. The 147 \HI\ sources that are found to be multi-galaxy systems are excluded from Table \ref{tab:param_desc}. 

\begin{figure}[htb!]
	\center
\includegraphics[width=3in,keepaspectratio]{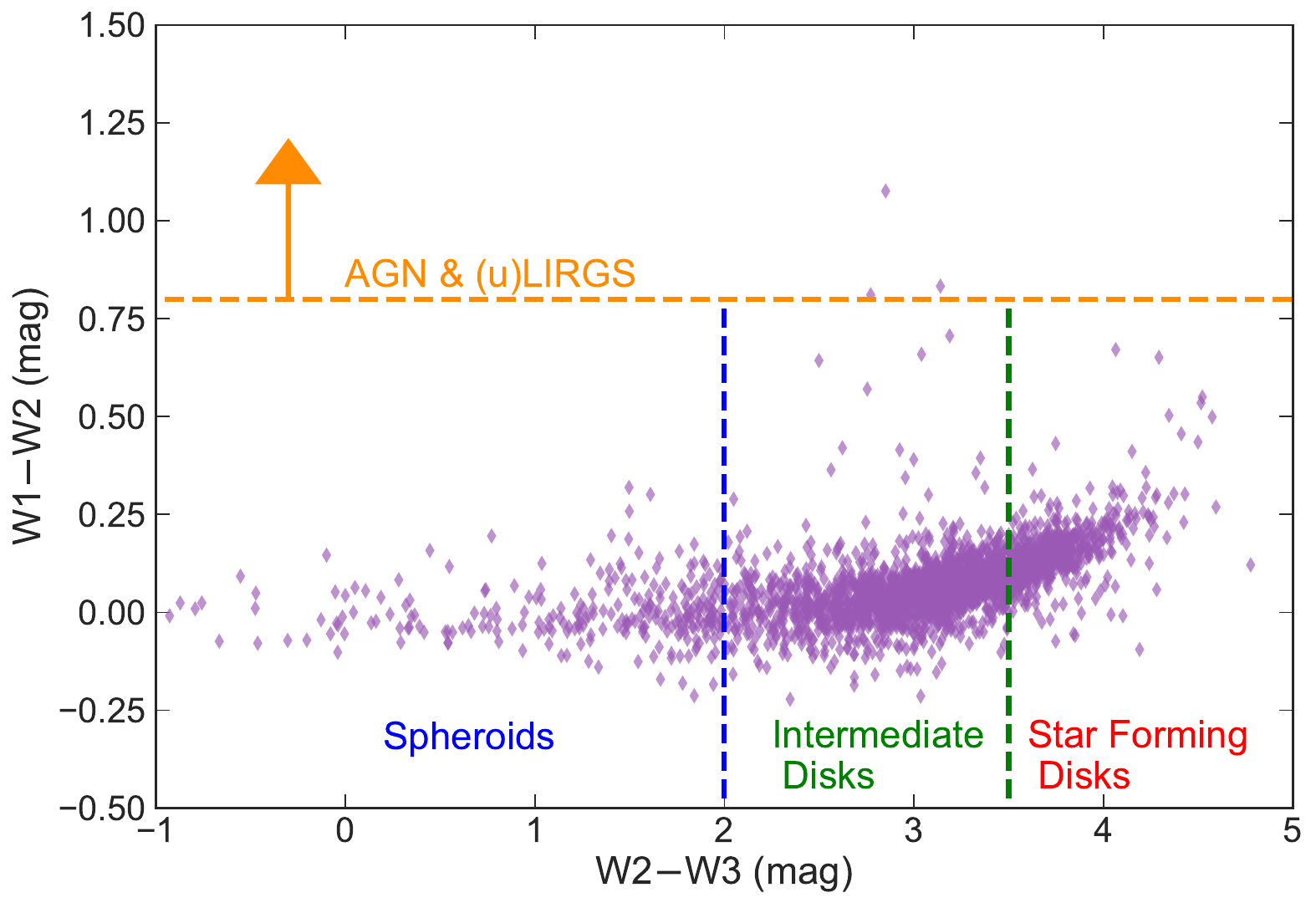}
	\caption{WISE mid-infrared colors of the HICAT sources with good WISE photometry. The horizontal and vertical lines denote the division between different types of galaxies \citep[e.g.,][]{jarrett17}. HICAT is dominated by star-forming, intermediate or late-type disk galaxies. }
	\label{fig:colorcolorPlot}
\end{figure}

Figure \ref{fig:colorcolorPlot} illustrates the WISE colors of HICAT galaxies, along with the expected colors of different types of galaxies, and clearly demonstrates that HIPASS is dominated by star-forming spiral galaxies, with relatively few ellipticals and luminous infrared galaxies (LIRGs). The galaxies with intermediate disk colors in AGN/LIRGs region may harbor dust-obscured AGNs and Seyferts \citep{jarrett11,huang2017}.

\section{Stellar Mass, \HI\ Mass and Star Formation Rate Estimation}\label{measurable}

In this section, we will describe the methods used to estimate stellar mass, \HI\ mass and star formation rate (SFR) for our samples. For these quantities, we compiled distances from (in order of preference) Cosmicflows-3 \citep{tully16} and HICAT. When distances are not available in the Cosmicflows-3 database, we determine luminosity distances using HIPASS redshifts by applying a Cosmic Microwave Background (CMB)-frame correction using the CMB dipole model presented by \citet{fixsen96}, followed by an application of the LCDM. For samples selected by stellar mass (Sections \ref{spiral_des} and \ref{Mss_des}), we include some galaxies that are not in HICAT, and for these galaxies, we use luminosity distances from velocities reported by NED in the same manner as stated above when distances are not available in the Cosmicflows-3 database. Table \ref{tab:param_desc} lists these distances---and their respective sources---for each galaxy.

\subsection{Stellar Mass}
The W1 and W2 bands are dominated by light from K- and M-type giant stars, and thus trace the continuum emission from evolved stars with minimal extinction at low redshifts. Consequently, these bands are good tracers of the underlying stellar mass of a galaxy \citep{meidt12,cluver14}. However, the W2 band is also sensitive to hot dust, as well as 3.3 $\upmu$m polycyclic aromatic hydrocarbon (PAH) emission from extreme star formation and active galactic nuclei (AGN). W1 contains the 3.3 $\upmu$m PAH emission, but typically weak for normal star-forming galaxies \citep{ponomareva18}. As such, the aggregate stellar mass will be overestimated in the presence of an AGN \citep{jarrett11, stern12, meidt14}. In order to mitigate this problem, we exclude AGNs from our analysis that are identified in the AGN catalog of \citet{veron10}.

Stellar masses were estimated following the GAMA-derived stellar mass-to-light ratio ($M_\ast/ L_{\mathrm{W1}_{\mathrm{Sun}}}$) relation of \citet{cluver14}:
\begin{eqnarray} \label{eq:GAMA}
\log_{10}M_\ast/ L_{\mathrm{W1}_{\mathrm{Sun}}}= -1.96(\mathrm{W1} - \mathrm{W2}) - 0.03,
\end{eqnarray}
which depends on the ``in-band" luminosity relative to the Sun,
\begin{eqnarray}
L_{\mathrm{W1}_{\mathrm{Sun}}} = 10^{-0.4(M - M_{\mathrm{Sun}}) },
\end{eqnarray}
where $M$ is the absolute W1 magnitude and $M_{\mathrm{Sun}}$ $=$ 3.24 \citep{jarrett13}. 
Equation \eqref{eq:GAMA} was determined using Galaxy and Mass \citep[GAMA;][]{drive09} survey with stellar masses derived assuming a \citet{chabrier03} IMF \citep{taylor11, cluver14}, and limited to galaxies with -0.05 $<$ W1$-$W2 $<$ 0.30, so we restrict the input colors for Equation \eqref{eq:GAMA} accordingly. The W1-W2 color dependence takes into account morphological dependence on the M/L and other factors, such as metallicity \citep{cluver14}. To determine the W1-W2 color, apertures are matched between the two bands. We typically use the W2 elliptical isophotal aperture as the fiducial since it is less sensitive than W1 and usually 10 to 15\% smaller in radial extent. This is particularly the case for galaxies whose mid-infrared emission is dominated by stellar light \citep[see][]{cluver17, jarrett17}.
 
The consistency of our stellar masses and those from other studies can be tested using MPA-JHU catalog \citep{brinchmann04}\footnote{Available via http://www.mpa-garching.mpg.de/SDSS/} for the Sloan Digital Sky Survey \citep[SDSS;][]{yor00}. By construction, our WISE stellar masses agree with the SED stellar masses determined by GAMA \citep{drive09,taylor11}, with a $1\sigma$ scatter of just 0.2 dex, and the GAMA \Ms$/L$ agree with the MPA-JHU \Ms$/L$, with a bi-weight mean and $1\sigma$ scatter in the difference between \Ms$/L$ of −0.01 and 0.07 dex. For dwarf galaxies the difference MPA-JHU between and UV-optical SED stellar masses \citep{huang12} is zero with the objects lying within $\pm$0.35 dex \Msolar \citep[][ and private correspondence]{maddox15}. Thus, we do not expect large offsets between the stellar masses we have obtained from WISE and those which have been used by recent studies of \HI\ galaxies.

\subsection{HI Mass}
To calculate the \HI\ mass, we use the published integrated 21 cm flux ($F_{\textup{H{\textsc i}}}$) as follows:
\begin{eqnarray}
\textup{M}_{\textup{H{\textsc i}}} [\textup{M}_\sun] = \frac{2.356\times10^5}{1+z} \times D_L^2\times \textup{F}_{\textup{H{\textsc i}}}[\textup{Jy km s}^{-1}],
\end{eqnarray}
where $D_L$ is the luminosity distance to the galaxy in Mpc and $z$ is the redshift measured from the \HI\ spectrum \citep[e.g.,][]{lutz17}. The uncertainty in the \HI\ mass ($\Delta$\Mh) is estimated using the method suggested by \citet{doyle06}:
\begin{align}
\Delta F_{\textup{H{\textsc i}}} &= 0.5\times F_{\textup{H{\textsc i}}}^{1/2}, \\
\Delta \textup{M}_{\textup{H{\textsc i}}} &= \textup{M}_{\textup{H{\textsc i}}}  \frac{\Delta F_{\textup{H{\textsc i}}}}{F_{\textup{H{\textsc i}}}}.
\end{align}

\subsection{Star Formation Rate}

The W3 and W4 bands are sensitive to the interstellar medium, active galactic nuclei and star formation \citep[e.g.,][]{calzetti07,jarrett11,cluver14,cluver17}. W4 emission is dominated by warm dust, and for star-forming galaxies W4 luminosity can be used to predict Balmer decrement correct H$\alpha$ luminosity with an accuracy of 0.2 dex \citep{brown17}. However, W4 lacks sensitivity and 39\% of HICAT sources lack W4 detections. 

WISE W3 luminosity includes contributions from PAHs, nebular emission lines, silicate absorption and warm dust, all of which are associated with star formation in galaxies. For $\sim L^\ast$ star-forming galaxies, \citet{cluver17} find PAHs and warm dust make 34\% and 62.5\% contributions, respectively, to the observed W3 luminosity. That said, we expect the contribution of PAHs to the W3 luminosity to decrease with decreasing galaxy mass due to the mass-metallicity relation of galaxies. Also, W3 better predicts the total infrared luminosity (and hence SFR) than W4 \citep{cluver17}. WISE W3 is thus a good star formation rate indicator, can be used to Balmer decrement corrected Ha luminosity with an accuracy of 0.28 dex \citep[i.e.,][]{brown17}.

Although the W3 band traces emission from star formation, it may also have contributions from evolved stellar populations. For $\sim L^\ast$ galaxies located at the centre of the star-forming main sequence, stellar continuum contributes 15.8\% of the W3 light and we subtract from our data using the W1 photometry and the method of \citet{helou04}. To account for this, we therefore scale the W1 integrated flux density, and subtract it from the W3 total flux to give an estimate of the W3 emission from the ISM, W3$_{\mathrm{PAH}}$ \citep{cluver17}. We use the prescription in Table 4 from \citet{brown17} to estimate the Balmer decrement corrected H$\alpha$ ($L_{H_{\alpha},\mathrm{Corr}}$):
\begin{align}
\nonumber \log{L_{W3_{\mathrm{PAH}}}}\; [\textup{erg s}^{-1}]  = &(40.79 \pm 0.06) + (1.27 \pm 0.04) \times \\ &(\log{L_{H_{\alpha},\mathrm{Corr}}} \;[\textup{erg s}^{-1}] -40),
\end{align}
with a $1\sigma$ scatter of 0.28 dex. SFRs are estimated by scaling the \citet{kennicutt98} calibration to a \citet{chabrier03} IMF:
\begin{equation}
SFR\; [M_\sun\: \textup{yr}^{-1}]=( 4.6\times10^{-42} ) \times L_{H_{\alpha},\mathrm{Corr}}\;[\textup{erg s}^{-1}] .
\end{equation}
The uncertainty in SFR is dominated by the scatter in the relationship between WISE W3 luminosity and Balmer decrement corrected H$\alpha$ luminosity and therefore the uncertainty in log(SFR) $\sim$ 0.28 dex \citep{brown17}.

We use the SFR calibration from \citet{brown17} because it provides better SFR estimates for a broad range of galaxies---including LIRGs and blue compact dwarfs galaxies---compared to the prior literature. Also, \citet{cluver17} compared the SFR calibrations from the prior literature to their calibration derived from total infrared luminosity and found the SFR calibration from \citet{brown17} to agree with their own.

\section{Samples}\label{samples}

We create three samples to address specific science questions: an \HI\ selected sample (\HI\ sample), a stellar mass-selected sample (\Mss) and a \spiral. In addition to addressing specific science questions, these samples allow us to compare to the prior literature and to explore the impact that galaxy morphology and selection bias have on the scaling relationships between star formation, stellar mass, and \HI\ mass.

For the remainder of the paper we focus on galaxies (including 3,513 HICAT galaxies) that are at least 10 degrees away from the Galactic plane and are not known AGNs (from \citet{veron10} catalog), although we do include some additional galaxies in our summary of galaxy coordinates, redshifts, and photometry provided in Table \ref{tab:param_desc}. We also exclude objects north of the main HICAT footprint ($\delta$ $\ge$ $+2$) from the \Mss\ and \spiral.

\subsection{\HI\ Selected Sample} \label{HIsample}
\begin{figure}[htb!]
	\center
\includegraphics[width=3in,keepaspectratio]{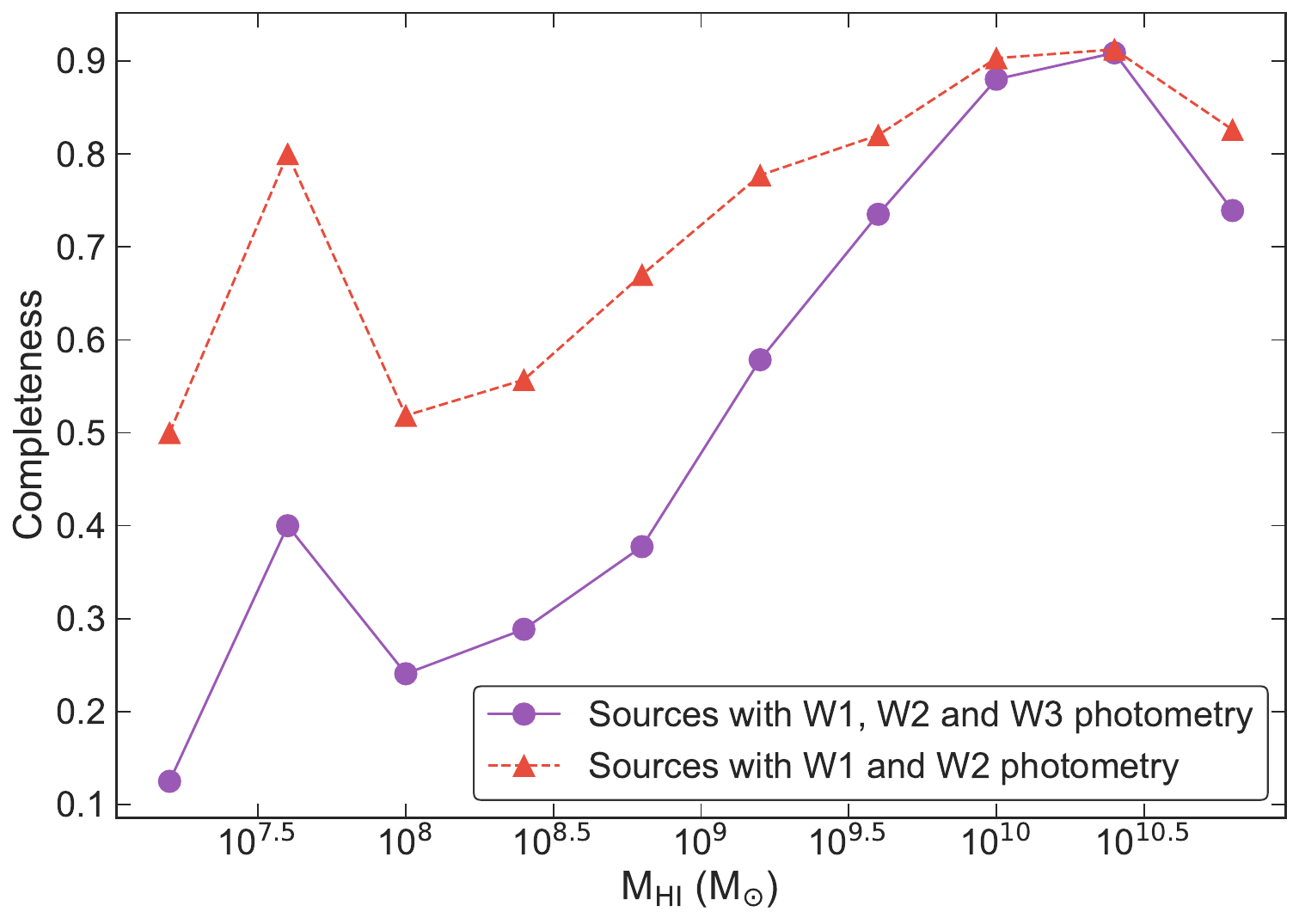}
	\caption{The fraction of HICAT sources with WISE photometry, compared to the total sample, as a function of \HI\ mass. Out of 3,513 HICAT sources, 80\% have good W1 and W2 detections and 68\% have good W3 detections.}
	\label{fig:HiComp}
\end{figure}

We use the \HI\ Parkes All-Sky Survey catalog (HICAT) to form the basis of the \HI\ selected sample. Out of 3,513 HICAT sources, 2,826 have good W1 and W2 photometry. Of these, 2,396 also have good photometry for the W3 band, and 2,342 galaxies in this category have significant W3$_{PAH}$ flux (W3 flux with the stellar continuum subtracted). Figure~\ref{fig:HiComp} shows the percentage of \HI\ sources with WISE counterparts as a function of \HI\ mass. We achieve a HICAT\--WISE match completeness of 80\% for \HI\ mass $>$ 10$^{9.5}$ \Msolar. 

\subsection{Spiral Sample}\label{spiral_des}

To measure the distribution of \HI\ masses and SFRs as a function of stellar mass and morphology, we generate a stellar mass-selected sample of spiral galaxies (the \spiral). The \spiral\ is drawn from the \citet{bonne15} catalog, which achieves a 99\% completeness in redshifts and morphologies for galaxies with $K_{\mathrm{tot}}$ $<$ 10.75. We maximize \HI\ completeness by limiting the sample to spiral galaxies (defined with de Vaucouleurs T-type $\ge$ 0) with redshift $\le$ 0.01 and re-measured W1 magnitude $<$ 10. Our \spiral\ consists of 600 galaxies.

\begin{figure}[htb!]
	\center
	\includegraphics[width=3in,keepaspectratio]{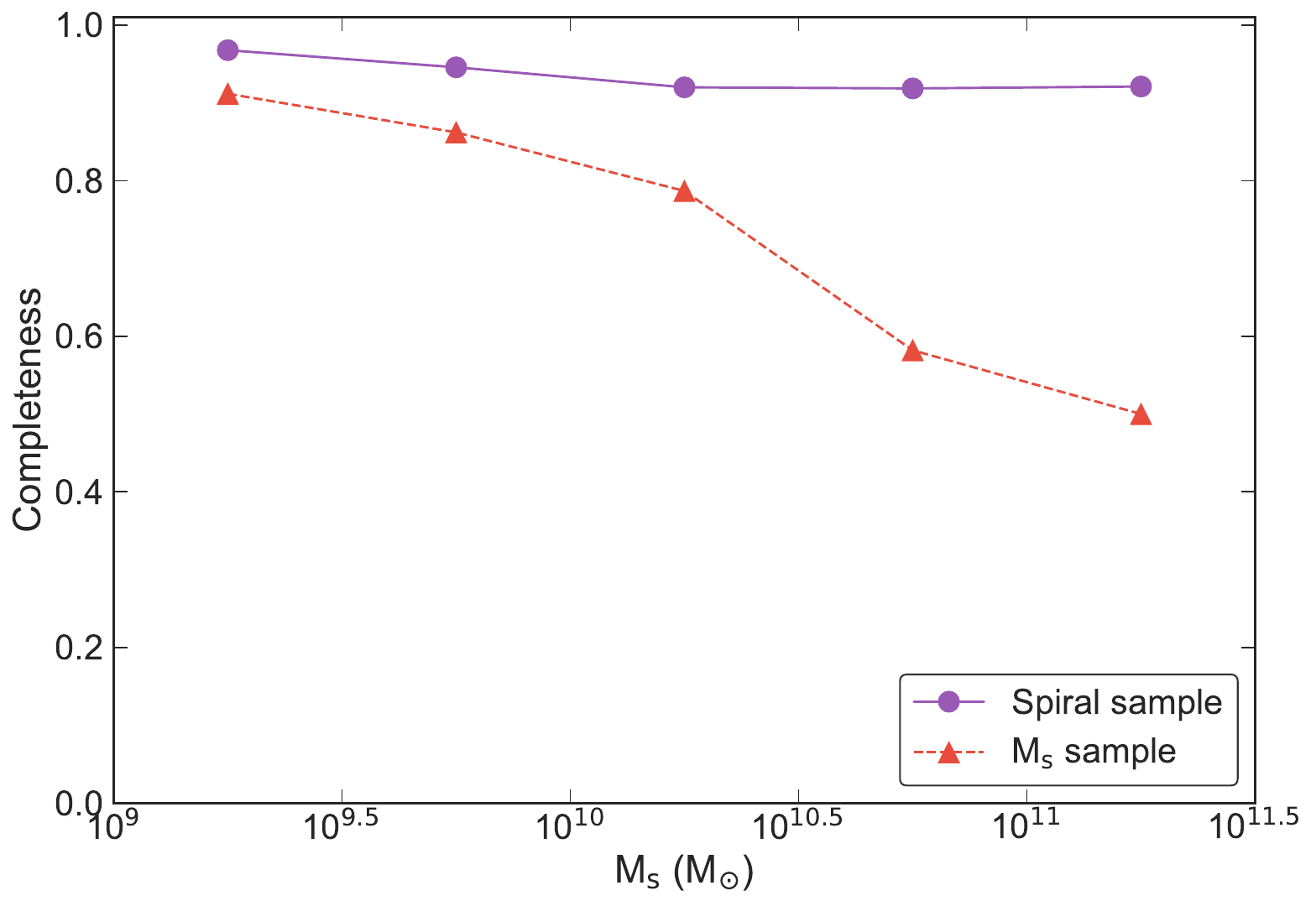}
	\caption{The fraction of galaxies in the \spiral\ and the \Mss\ with \HI\ mass measurements. For the spiral sample, 556 out of 600 galaxies have a \HI\ counterpart, giving the \spiral\ a completeness of 93\%. 590 galaxies out of the 839 galaxies in the \Mss\ have a \HI\ counterpart. }
	\label{fig:MsComp}
\end{figure}

For 435 of our spiral galaxies we obtained \HI\ fluxes from HICAT, while for a further 121 galaxies we obtained archival \HI\ fluxes from \citet{paturel03}, \citet{huchtmeier89}, \citet{springob05} and \citet{masters14}. The details of the \HI\ counterparts are provided in Table \ref{table:Ms}. As we illustrate in Figure \ref{fig:MsComp}, 93\% of the spiral galaxies have \HI\ detections. 

\begin{table}[ht]
\centering
\caption{The number of galaxies the \spiral\ and the \Mss\ with \HI\ counterparts. }
\label{table:Ms}
\begin{tabular}{@{}ccc@{}}
\toprule
\HI\ Source & Spiral Sample & \Mss \\ \hline 
Total & 600 & 839 \\
HICAT & 435 & 454 \\
\citet{paturel03} & 100 & 114 \\
\citet{huchtmeier89} & 13 & 13 \\
\citet{springob05} & 1 & 2 \\
\citet{masters14} & 7 & 7 \\ \hline 
\end{tabular}
\end{table}

The aforementioned W1 and redshift limits are chosen because of the brightness limitation of the parent sample and the detection sensitivity of HICAT. While the notional limit for the \citet{bonne15} is $K_{tot}$ $\le$ 10.75, this limits the $W1$ $\le$ 11.5 and galaxy numbers decline at $W1$ $>$10. Table \ref{table:cuts} lists the number of galaxies in our sample, the number with HI detections and the percentage with HI detections with and without the redshift and W1 magnitude limits applied. Removing the redshift and magnitude limits from the spiral sample decreases the percentage of galaxies with HI detections to 50\%, but has little impact on measured relations we describe in Section \ref{MassRelations}.
\begin{table*}[htb!]
\centering
\caption{The basic properties of our \spiral\ and spiral galaxy samples without the redshift and W1 selection criteria applied.}
\label{table:cuts}
\begin{tabular}{lcccc}
\hline\hline
& Spiral Sample & No W1-mag cut & No redshift cut & No W1 magnitude and redshift cuts applied \\ \hline
Total $\#$ of galaxies       & 600           & 792                 & 1442            & 3458            \\
$\#$ \HI\ counterparts  & 556           & 672                 & 1048             & 1697             \\
\% with HI counterpart & 93\%        & 87\%             & 73\%          & 50\%        \\  \hline
\end{tabular}
\end{table*}

\subsection{Stellar Mass-selected Sample (\Mss)}\label{Mss_des}
In order to compare the \HI\ mass, stellar mass and SFR relationships of this work to those of GASS \citep{catinella10,catinella12,catinella13}, which uses a stellar mass-selected sample, we have also produced such a sample (\Mss). The GASS sample is designed to measure the neutral hydrogen content of 1000 galaxies to investigate the physical mechanisms that regulate how cold gas responds to different physical conditions in the galaxy and the processes responsible for the transition between star-forming spirals and passive ellipticals \citep{catinella10}. 

The \Mss\ selection is identical to that of the spiral sample except that it lacks the T-type criterion. The \Mss\ contains 839 galaxies, of which 590 have an \HI\ counterpart. The details of the \HI\ counterparts are provided in Table \ref{table:Ms}. Figure \ref{fig:MsComp} shows that the fraction of galaxies with an \HI\ measurement drops at the higher stellar masses, where the number of \HI\--poor ellipticals increases.

\section{ \Mh\ - \Ms\ relationship} \label{MassRelations}

In this section, we will look at the relationship between \HI\ mass and stellar mass for all three samples. The \HI\---stellar mass relation is one of the principal means used to provide insight to the history of gas accretion and star formation. Also, as discussed in Section \ref{Discussion}, it can be used to test models of the stability of \HI\ disks and how these disks can fuel star formation \citep{wong16,obreschkow16}. For each sample, we bin the data into stellar mass bins with a width of $\log$(\Ms) $=$ 0.5 and for bins with $\ge$ 10 galaxies we measure the median \HI\ mass and $1\sigma$ scatter in \HI\ mass for each stellar mass bin. The $1\sigma$ scatter about the median is determined using the range encompassing 68\% of the data. For the \spiral\ and \Mss, care is needed when accounting for the galaxies with \HI\ non-detections. When estimating the median \HI\ mass as a function of the stellar mass, we assume the \HI\ upper limits are below the median \HI\ mass for the relevant stellar mass bin, which is a reasonable approximation when the \HI\ detection rate is $\gg$ 50\%. The median \HI\ masses are determined for any stellar mass bin with 10 or more galaxies and \HI\ detection rate above 50\%. Upper limits for individual galaxies are determined using the integrated flux of 7.5 Jy km s$^{-1}$, corresponding to the HICAT's 95\% completeness limit \citep{zwaan04}.

While we list the individual uncertainties in Table \ref{tab:param_desc}, we find that W1 $\sim$ 12 galaxies have a stellar mass uncertainty $\le$ 0.2 dex, and this uncertainty decreases with increasing W1 flux. Also, 80\% of the \HI\ sample has an \HI\ mass uncertainty better than 20\%.  

\subsection{HI Sample}
\begin{figure}[htb!]
		\includegraphics[width=3.3in,keepaspectratio]{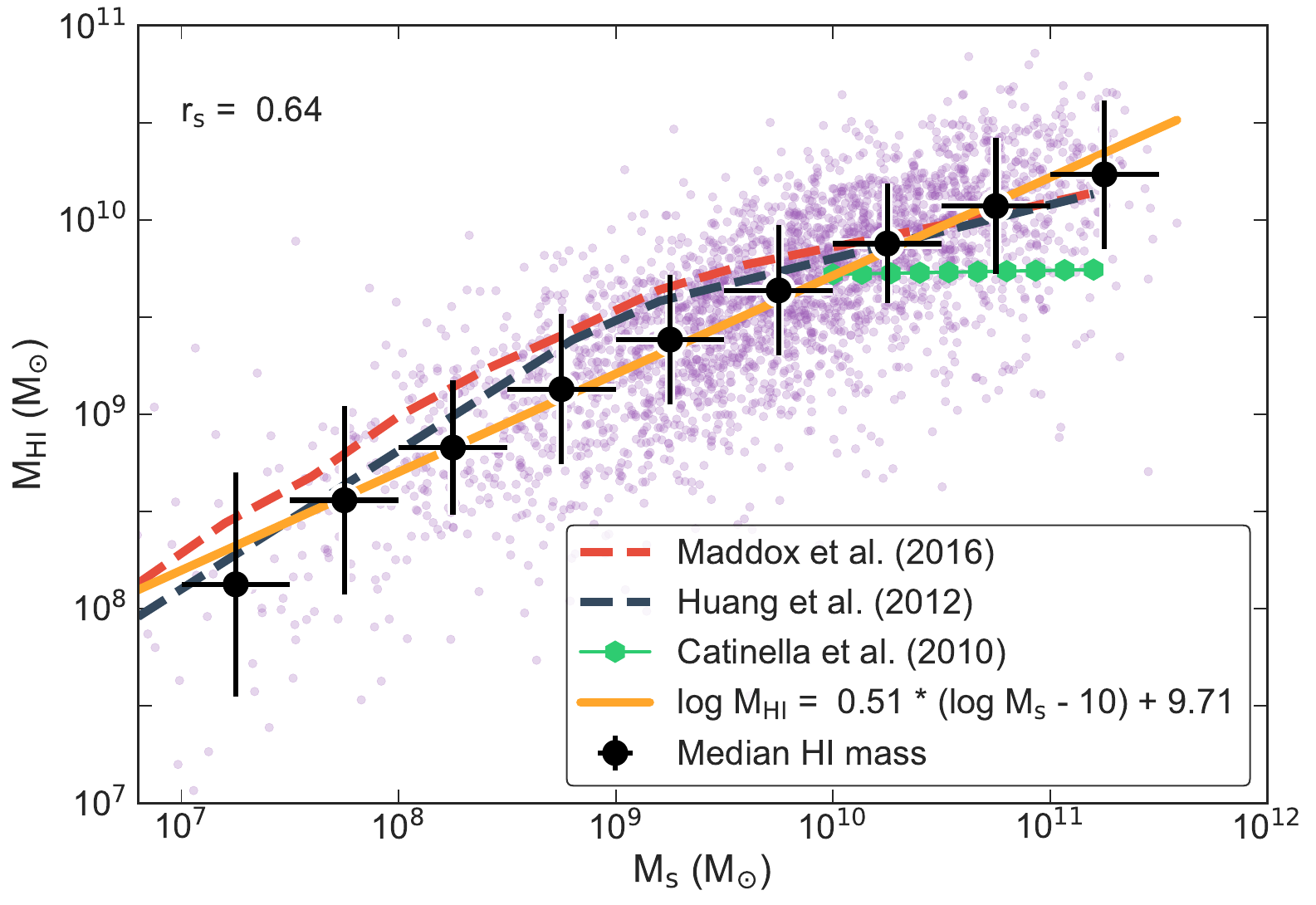}
	\caption{ \HI\ mass versus stellar mass of the \HI\ selected sample. Spearman's rank correlation, $r_s$, is listed in the top left corner. Our median \HI\ masses are slightly below those derived by \citet[blue dashed-line;][]{huang12}, and \citet[red dashed-line;][]{maddox15}, who both used ALFALFA \HI\ selected samples. However the \HI\ mass medians of the \HI\ sample are higher than the GASS sample \citep[green solid-line;][]{catinella10}, which is a stellar mass-selected sample. \HI\ samples overestimate \HI\ mass as a function of stellar mass because \HI\ samples do not detect galaxies with low \HI\ mass, such as ellipticals. }
	\label{fig: MhiMs_Mhi}
\end{figure}

The relationship between \HI\ mass and stellar mass for the \HI\ selected sample is illustrated in Figure \ref{fig: MhiMs_Mhi}. The \HI\ mass is a strong function of stellar mass among the \HI\ sample (Spearman's rank correlation, $r_s$, $=$ 0.64) and the least-squares fit to the medians, represented by an orange line in Figure \ref{fig: MhiMs_Mhi}, is  
\begin{equation}
\log{\; M_{\textup{H{\sc i}}}} = 0.51\:(\log{\;M_\ast} - 10) + 9.71.
\end{equation}
We fit 68\% of the \HI\ masses are within 0.5 dex of our best-fit relation.

\HI\ mass versus stellar mass relations for the \HI\ selected sample and previous studies \citep{catinella10,huang12,maddox15} are also plotted in Figure \ref{fig: MhiMs_Mhi}. Our sample and the ALFALFA samples of \citet{huang12} and \citep{maddox15} are \HI\ selected, while the GASS sample of \citet{catinella10} is stellar mass-selected. We find the relations for the \HI\ selected samples are qualitatively similar, with median \HI\ mass increasing with stellar mass. In contrast, the \HI\ versus stellar mass relation measured with the GASS sample \citep{catinella10} is up to 0.5 dex lower than those derived from \HI\ selected samples, as the GASS sample includes galaxies with low \HI\ masses (including ellipticals). There are also discrepancies in the \HI\ versus stellar mass relations measured with different \HI\ selected samples. We do not see the break in the relation at a stellar mass of 10$^{9}$ \Msolar\ that was previously observed by ALFALFA \citep[i.e.,][]{huang12,maddox15}\footnote{The slopes of the least-squared fits for Figure \ref{fig: MhiMs_Mhi} is 0.65$\pm$0.014 for stellar masses $\le$ 10$^9$ \Msolar\ and 0.48$\pm$0.013 for stellar mass $>$ 10$^9$ \Msolar\ and therefore are consistent with each other. Meanwhile, \citep{huang12} measured a slope of 0.712 for stellar masses $\le$ 10$^9$ \Msolar\ and 0.276 for stellar masses $>$ 10$^9$ \Msolar}. Below a mass of 10$^{9}$ \Msolar\ our sample is less than 70\% complete for WISE counterparts, and thus we may not be reliably measuring HI mass versus stellar mass in this mass range However, even if this was not an issue we believe this sample would produce a biased relation, as it (by construction) excludes galaxies that have high stellar masses but low HI masses (i.e., many elliptical galaxies).

\subsection{Spiral Sample}
\begin{figure}[h!]
	\center		\includegraphics[width=3.3in,keepaspectratio]{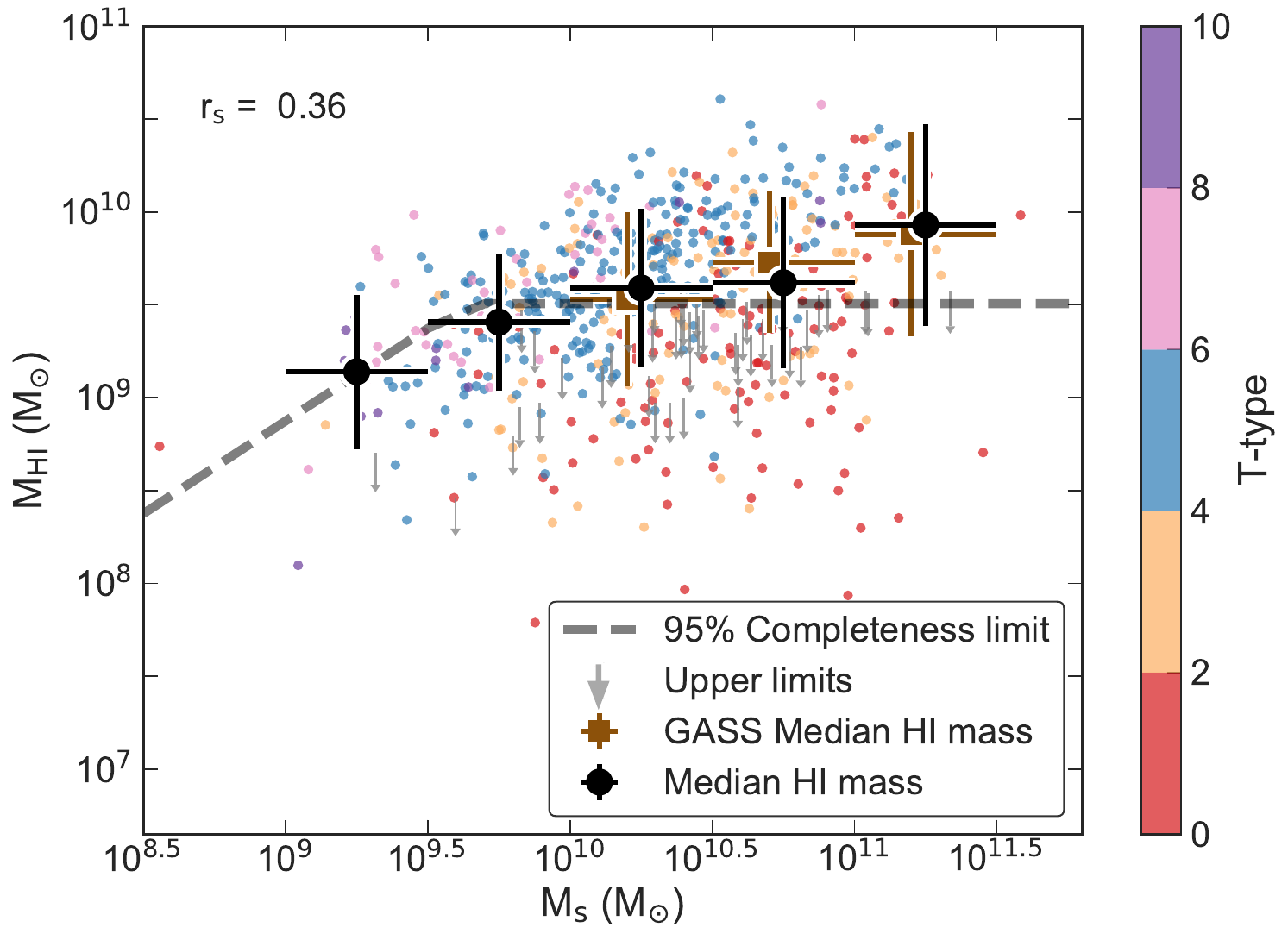}
	\caption{\HI\ mass versus stellar mass for the (highly complete) \spiral. Symbols are color coded by morphology, the median HI masses are shown for the \spiral\ (black circles) and the GASS sample \citep[brown squares;][]{catinella10}. The dashed line shows the \HI\ mass limit for $z$ $\le$ 0.01 and $W1$ $\le$ 10 star-forming galaxies. The Spearman's rank correlation coefficient is shown in the the top left. \HI\ mass is a function of both T-type and stellar mass. The median \HI\ masses increases with stellar mass with a power-law index of 0.35. }
	\label{fig:MhiMs_spiral}
\end{figure}

The \HI\ and stellar mass distribution for the spiral sample is shown in Figure \ref{fig:MhiMs_spiral}. The median \HI\ mass increases with stellar mass with a least-squared fit of
\begin{equation}
\log{\; M_{\textup{H{\sc i}}}} = 0.35\:(\log{\;M_\ast} - 10) + 9.45,
\end{equation}
with 68\% of the \HI\ masses within 0.4 dex.

However, as Figure \ref{fig:MhiMs_spiral} illustrates, at a given stellar mass the median \HI\ mass increases with T-Type while the dispersion decreases with T-Type. For example, for the 10$^{10}$ to 10$^{10.5}$ \Msolar\ stellar mass bin, the median \HI\ mass and the $1\sigma$ spread of galaxies for all spirals is 10$^{9.53}$ \Msolar\ $\pm$ 0.47 dex, for T-type 0 to 2 is 10$^{9.26}$ \Msolar\ $\pm$ 0.59 dex, and for T-type 6 to 8 is 10$^{9.72}$ \Msolar\ $\pm$ 0.31 dex. The increasing spread of HI masses with decreasing T-type for spiral galaxies may be part of a broader trend, as \citet{serra12} concluded that the HI mass distribution for early-type galaxies was far broader than that for spirals. They suggest this scatter reflects the large variety of \HI\ content of early-type galaxies, and confirms the lack of correlation between \HI\ mass and luminosity.

In Figure \ref{fig:MhiMs_spiral}, we compare our \HI\ mass-stellar mass distribution of our \spiral\ to that from GASS \citep{catinella10}, using GASS galaxies that we have classified as spirals with Galaxy Zoo 1 \citep[GZ1;][]{lintott11}. Using the GZ1 classifications and a 70\% vote threshold, we find GASS comprises 305 spirals, 273 elliptical and 182 galaxies with uncertain morphology\footnote{Using the default criteria of 80\% vote threshold \citep[see the following references for details on Galaxy Zoo and the data release:][]{lintott08,lintott11}, 291 galaxies (39\%) in GASS are classified as unknowns. We have decreased the vote requirement to 70\% to decrease the number of unknowns to 182 galaxies (24\%), although we find this has little impact on our measured relations.}. We repeat our analysis on the GASS sample, using the same stellar mass bins and \HI\ median mass calculations with stellar mass bins with 10 or more galaxies and an \HI\ detection rate $>$ 50\%. The estimated median \HI\ masses for the spiral sample, and the GASS spiral samples are listed in Table \ref{table:spiral_med}. The median \HI\ masses for the GASS spiral sample increases with stellar mass similar to our \spiral. The median \HI\ mass of the spiral samples differs about 0.08 dex on average. To explain the relationship between \HI\ mass and stellar mass of the spiral samples we turn to the halo spin parameter models of \citet{obreschkow16} in Section \ref{Discussion}.

\begin{table*}[htb!]
\center
\caption{Median \HI\ mass and $1\sigma$ (68\%) scatter as a function of stellar mass for the \spiral, and the GASS spiral sample.}
\label{table:spiral_med}
\begin{tabular}{|c|cccc|cccc|}
\toprule
& \multicolumn{4}{c|}{Spiral Sample}&\multicolumn{4}{c|}{GASS Spiral Sample}\\ 
log(\Ms) & log(\Mh) & $1\sigma$ & Total& \% \HI\    &   log(\Mh) & $1\sigma$& Total & \% \HI\ \\ 
(\Msolar) & (\Msolar)  & (\Msolar) & & Detections& (\Msolar)  & (\Msolar) & &Detections  \\\hline
9.25   & 9.14    & 0.42   & 31  &97  &&&&   \\
9.75     & 9.41     & 0.37  & 129 &95   &&&&   \\
10.25    & 9.59     & 0.43   & 225  &92   &  9.53 &0.47  &130  & 89   \\
10.75     & 9.61     & 0.46   & 174  &97   &9.73&0.38&111&94   \\
11.25    & 9.93     & 0.54   & 38 &92 &9.88&0.55&64&83   \\ \hline
\end{tabular}
\tablecomments{Median \HI\ masses are not calculated for stellar mass bins with an \HI\ completeness $\le$ 50\%. }
\end{table*}

\subsection{M$_s$ Sample}
\begin{figure}[htb!]
	\center
\includegraphics[width=3.3in,keepaspectratio]{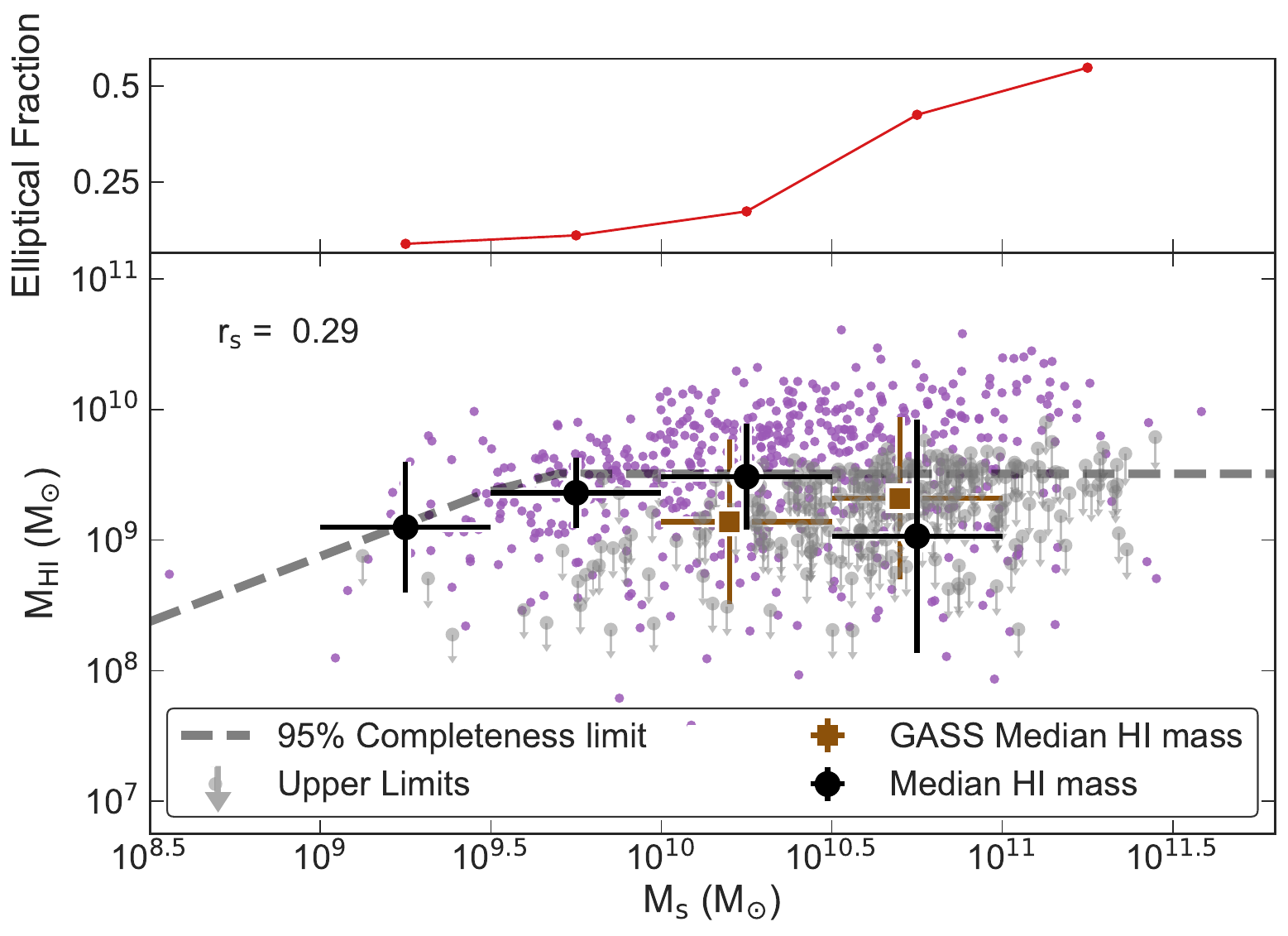}
	\caption{The spiral fraction versus stellar mass (upper panel) and \HI\ mass versus stellar mass (lower panel) for the \Mss. Median HI masses are shown for the \Mss\ (black circles) and the GASS sample \citep[brown squares;][]{catinella10}. The HIPASS mass limit for galaxies at $z$ $\le$ 0.01 and $W1$ $\le$ 10 is shown by the gray dashed line and the Spearman's rank correlation coefficient is shown in the top left. \HI\ mass is almost flat with increasing stellar mass for both samples, in large part because of the increasing fraction of passive early-type galaxies with increasing stellar mass. }
	\label{fig:MhiMs_Ms_all}
\end{figure}

In Figure \ref{fig:MhiMs_Ms_all}, we present \HI\ mass versus stellar mass for our \Mss\ and the equivalent stellar mass-selected sample of GASS \citep{catinella10}. The estimated median \HI\ masses of the \Mss\ and the GASS sample are listed in Table \ref{table:Ms_med}. For the \Mss\ the \HI\ mass increases with the stellar mass for the stellar mass bins $\le$ 10$^{10.5}$ \Ms\ and then flattens for the highest stellar mass bin. The estimated \HI\ mass median for the highest stellar mass bin may be underestimated, as the \HI\ completeness for this bin is only 58\% and our assumption that all non-detections are below the median could be in error. 

At stellar masses greater than 10$^{10}$ \Msolar, median \HI\ mass is almost constant with stellar mass, and our measurements agree with those of GASS to within 0.3 dex. This is in contrast with the trend shown in Figure \ref{fig:MhiMs_spiral} for the \spiral. The obvious explanation, given the prior literature \citep[e.g.,][]{catinella10, huang12}, is that this is due to the increasing fraction of gas poor early-type galaxies at high \Ms, as illustrated in the top panel of Figure \ref{fig:MhiMs_Ms_all}. \citet{serra12} found that early-type galaxies host less \HI\ than spiral galaxies, but have a broader range of \HI\ masses. For example, \citet{serra12} find that elliptical galaxies have \HI\ mass from 10$^{7}$ to 10$^{9}$ \Msolar (the lower limit is uncertain as this overlooks \HI\ non-detections in their sample), while the \HI\ mass distribution for spirals peaks at $\sim$ 2$\times$ 10$^9$ \Msolar, with a small number of galaxies below 10$^8$ \Msolar. Combining this result with our previous findings from our spiral sample, we conclude that as one moves from early-type to late-type galaxies, median HI mass increases while the scatter in HI mass decreases. Within an individual T-type, \HI\ mass typically increases with stellar mass, and it is the increasing fraction of early-types with increasing stellar mass that explains the roughly constant median \HI\ masses measured for the \Mss. 

\begin{table*}[htb!]
\center
\caption{Median \HI\ mass and $1\sigma$ (68\%) scatter as a function of stellar mass for the \Mss, and the GASS sample. }
\label{table:Ms_med}
\begin{tabular}{|c|cccc|cccc|}
\toprule
& \multicolumn{4}{c|}{\Mss}&\multicolumn{4}{c|}{GASS Spiral Sample}\\ 
log(\Ms) & log(\Mh) & $1\sigma$ & Total&\% \HI\ &   log(\Mh) & $1\sigma$& Total & \% \HI\ \\ 
(\Msolar) & (\Msolar)  & (\Msolar) & & Detections& (\Msolar)  & (\Msolar) & &Detections  \\\hline
9.25 & 9.10& 0.50& 34  &91  &&&&   \\
9.75 & 9.56 & 0.27 & 145 &86   &&&&   \\
10.25 & 9.48 & 0.41 & 272  &  79 &  9.14 &0.64  &299 & 68   \\
10.75 & 9.03 & 0.89 & 299  &58   &9.32&0.62&292&63   \\
11.25 &&& 84 & 50 &&& 168 & 50 \\ \hline
\end{tabular}
\tablecomments{Median \HI\ masses are not calculated for stellar mass bins with an \HI\ completeness $\le$ 50\%.}
\end{table*}
 
\subsection{The Impact of Sample Selection}
\begin{figure}[htb!]
	\center		\includegraphics[width=3.3in,keepaspectratio]{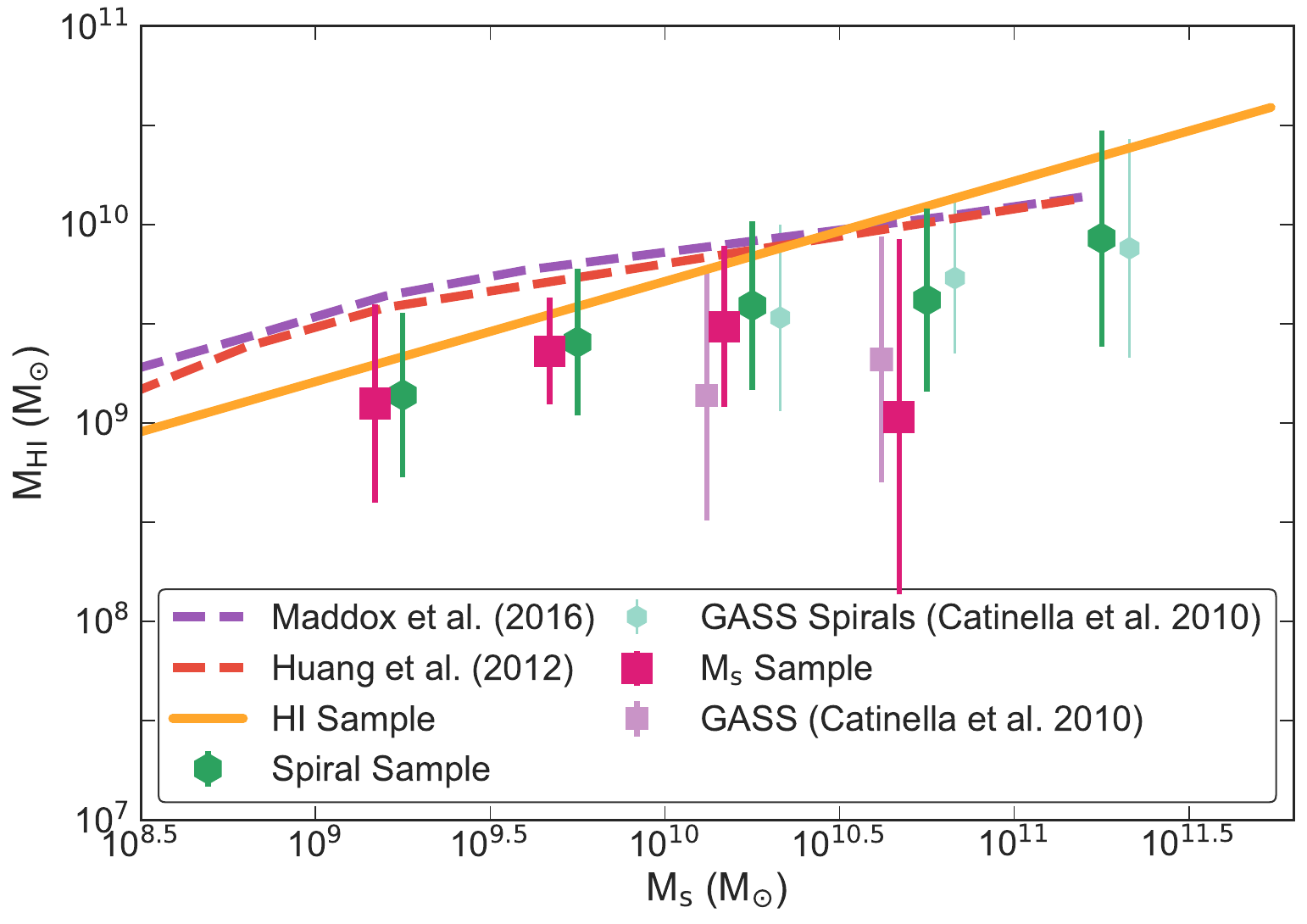}
	\caption{Comparison of HI mass versus stellar mass for our \HI\ selected sample, the \spiral, the \Mss\ (without morphological criteria) and the prior literature. \HI\ selected samples give consistently higher relations than the other samples, as (by definition) they exclude galaxies with comparatively low \HI\ masses.}
	\label{fig:MhiMs_Ms_trends}
\end{figure}

A key conclusion from the previous sections is sample selection impacts measured \HI\ mass versus stellar mass relations, and to illustrate this in Figure \ref{fig:MhiMs_Ms_trends} we plot \HI\ mass--stellar mass relations for \HI\ selected samples, spiral galaxy samples and stellar mass selected samples, including data from both our work and the literature. For all values of stellar mass, \HI\ selected samples have a higher \HI\ mass than spiral-selected and stellar mass-selected samples. This is because \HI\ surveys are designed to sample a large number of \HI\--rich systems, and therefore lack the sensitivity to detect the \HI\--poor galaxy population. For example, the HIPASS survey can detect galaxies with \HI\ masses $>$ 10$^{9}$ \Msolar\ at $z$ $=$ 0.01; however, elliptical galaxies have \HI\ masses $\le$ 10$^{9}$ \Msolar\ \citep{serra12}, and would thus be largely missing from HIPASS samples at these redshifts. Even late-type galaxies in Figure \ref{fig:MhiMs_spiral} have HI masses as low as 10$^{8}$ \Msolar, and thus some are missing from HIPASS selected samples at $z$ $>$ 0.01. Similar selection effects apply to ALFALFA, albeit at higher redshifts. This is not surprising and indeed was a motivation for studies such as GASS, but does illustrate that \HI\ mass versus stellar mass relations have a strong dependence on sample selection. 

\section{Star-Forming Properties of the Spiral Sample}\label{SFR}

\subsection{Star-forming Main Sequence}\label{MS}
While the relationship between SFR and \HI\ mass is the principal focus of the paper, we are also able to measure the local ($z \le 0.01$) star-forming main sequence \citep[MS; e.g.,][]{noeske07a, rodighiero11, wuyts11} using the \spiral. Measurements of the MS are enhanced by our highly complete sample, our new WISE photometry (which should mitigate aperture bias---see Section \ref{photometry}), and our ability to take advantage of a recent calibration of W3 as a SFR indicator that uses large aperture photometry \citep{brown17}.

\begin{figure*}[htb!] \includegraphics[width=\linewidth,keepaspectratio]{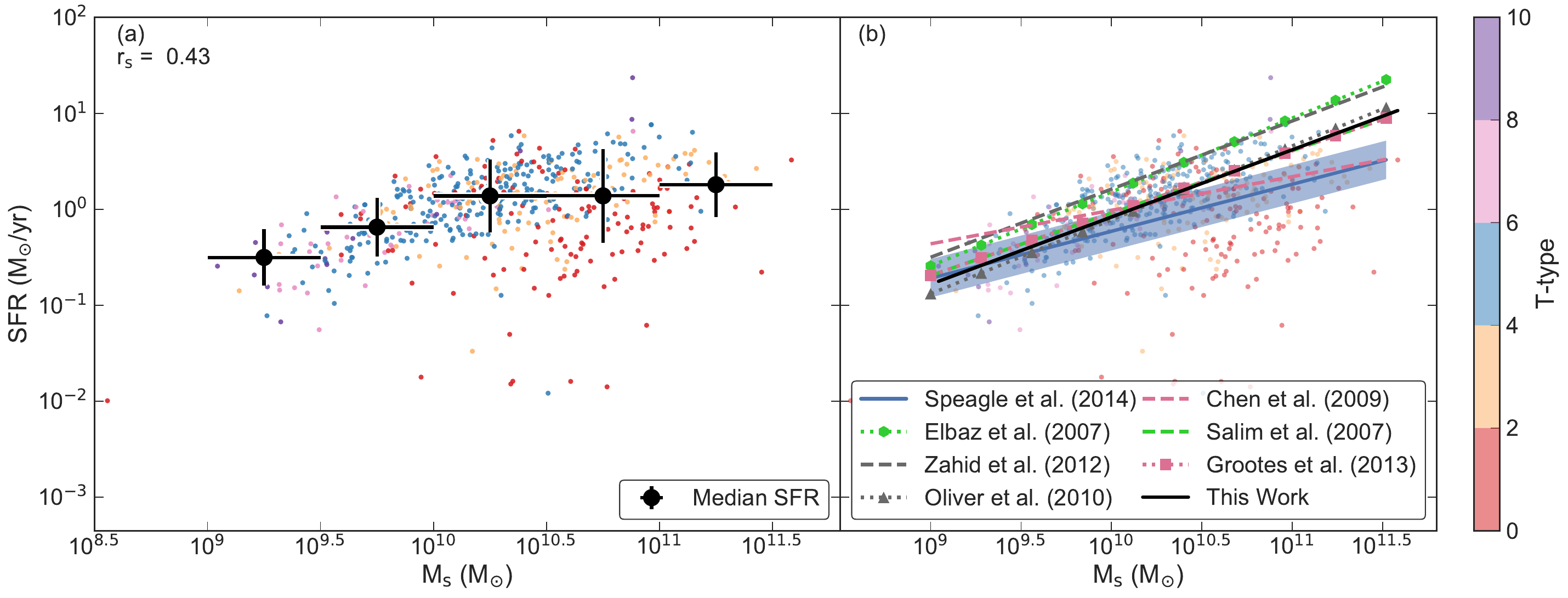}
\centering
\caption{SFR versus stellar mass for the \spiral\ with symbols color coded by morphology. The median SFRs are also shown in panel (a). The median SFR increases with stellar mass for \Ms\ $<$ 10$^{10.5}$ \Msolar\ and then flattens for \Ms\ $>$ $10^{10.5}$ \Msolar, indicating evidence of mass quenching. To mitigate the effect of mass quenching, we model the MS using spiral galaxies with \Ms\ $<$ $10^{10.5}$ \Msolar\ and find $\log$ $SFR$ $=$ 0.7 $\times$ ($\log$ \Ms $-$ 10) $-$ 0.09. As illustrated by panel (b), z$\sim$0.01 relations compiled by \citet{speagle14} show considerable scatter, and this may result (in part) from how studies exclude (or include) high mass spirals that may have already commenced quenching.}
	\label{fig:SFR}
\end{figure*}

In Figure \ref{fig:SFR} we present our star-forming main sequence. As expected, the median of the $\log$ $SFR$ increases from -0.50 dex for the $10^{9}$ to 10$^{9.5}$ \Msolar\ stellar mass bin to 0.14 dex for the 10$^{10}$ to 10$^{10.5}$ stellar mass bin, with the scatter of individual galaxies about the median being $\sim$ 0.3 dex. For stellar mass bins above 10$^{10}$ \Msolar, the median $\log$ $SFR$ is roughly constant at 0.14 while the scatter of the individual galaxy SFRs about the median increases from 0.38 to 0.49 dex. The changing trend of SFR with increasing stellar mass and the increased dispersion of SFRs is evidence of mass quenching \citep{kauff03}, and this also coincides with an increasing fraction of early-type spirals (T-type $\le$ 2). To mitigate the effect of mass-quenching on our model fit to MS, we only fit to galaxies with stellar mass $\le$ 10$^{10.5}$ \Msolar (shown in Figure \ref{fig:SFR}b) and measure the MS to be
\begin{equation}
\log{\; \mathrm{SFR}} = 0.7\:(\log{\;M_\ast} - 10) - 0.09
\end{equation}
with a $1\sigma$ scatter of 0.27 dex. Alternate selection criteria to mitigate the effect of mass-quenching produces similar MS fits. For example, a subsample of galaxies with T-type $>$ 2 produces a fit of $\log SFR = 0.61 (\log M_{\ast}-10) - 0.08$ ($1s\sigma$ scatter $=$ 0.26 dex).

Figure \ref{fig:SFR}b and Table \ref{table:MS} also compare the MS relation from this work to the prior literature \citep{zahid12,oliver10,chen09,elbaz07,salim07,grootes13}, with data taken from the extensive review by \citet{speagle14}. To shift the \citet{speagle14} homogenized MS relations from a Kroupa to Chabrier IMF, we apply a -0.03 and -0.07 dex shifts to the stellar masses and SFRs respectively. The blue line is the $z$ $=$ 0.01 MS relation given by Equation (28) from \citet{speagle14}, and the shaded region is the ``true" scatter about the MS \citep[for more details, please refer to][]{speagle14}. The normalization (at $\log$ \Ms\ $=$ 10) of our best fit is 0.04 dex smaller and the slope is 0.21 larger than the best fit for the MS of \citet{speagle14}. \citet{speagle14} note that the wide range for MS slopes for the local universe suggests that the systematics involved are underestimated, and they estimate the magnitude of these systematics on the MS slopes to be of the order of $\sim$0.2 dex. As we noted earlier, our slope does depend on the criterion used to reject galaxies that could be undergoing quenching, and including early-type spiral galaxies with masses above 10$^{10.5}$ \Msolar\ reduces our slope to 0.416 ($1\sigma$ scatter $=$ 0.34), which is closer to that of \citet{speagle14}.

\begin{table*}[ht]
\centering
\caption{Main-sequence relationships}
\label{table:MS}
\begin{tabular}{@{}llllllll@{}}
\toprule
Paper & $\alpha$ & $\beta$ & $\log$ $SFR(10)$ & $z_{\mathrm{med}}$ & $z_{\mathrm{range}}$ & log \Ms\ range & Survey\\ \hline 
This Study & 0.7 & -7.09 & -0.09& & $\le$ 0.01 & 9.0-11.0 & WISE \\
\citet{zahid12} & 0.71$\pm$0.01 & -6.78$\pm$0.1 & 0.32& 0.07 &  0.04-0.1 & 8.5--10.4 & SDSS \\
\citet{oliver10} & 0.77$\pm$0.02 & -7.88$\pm$0.22 &-0.18& 0.1& 0.0-0.2 & 9.1--11.6 & SWIRE  \\
\citet{chen09} & 0.35$\pm$0.09 & -3.56$\pm$0.87 &-0.06& 0.11 & 0.005-0.22 & 9.0--12.0 & SDSS \\
\citet{elbaz07}& 0.77 & -7.44&0.26& 0.06& 0.015-0.1  & 9.1-11.2& SDSS  \\
\citet{salim07} & 0.65 & -6.33 &0.17 & 0.11&  0.005-0.22  & 9.0-11.1 & GALEX-SDSS selected \\
\citet{speagle14} & 0.49 & -5.13 & -0.03 & & & & \\
\citet{grootes13}\tablenotemark{1} & 0.550 & -5.520 & -0.02& & $\le$ 0.13 & 9.5-11& GAMA/Herschel-ALTAS \\
\citet{cluver17}\tablenotemark{2} & 1.05$\pm$0.09 & -10.40$\pm$0.88 & 0.09& & z $<$ 0.01($<$ 30 Mpc)& 7-11.5& SINGS/KINGFISH \\
\hline 
\end{tabular}
\tablecomments{Column 1: Reference. Columns 2 \& 3: MS slope $\alpha$ and normalization $\beta$ reported in Table 6 of \citet{speagle14}. These best fit parameters have been adjusted for IMF, cosmology, SPS model and emission line corrections. Column 4: $\log$ $SFR$ predicted by each MS relation at $\log$ \Ms\ $=$ 10. Column 5: The median redshift. Column 6: Redshift range. Column 7: Stellar mass range. Column 8: Survey Data.}
\tablenotetext{1}{The normalization have not adjusted by \citet{speagle14}. We do not apply shifts stellar masses and SFRs respectively as \citet{grootes13} makes use of \citet{chabrier03} IMF.}\tablenotetext{2}{The normalization have not adjusted for systematics by \citet{speagle14}. The MS trend of \citet{cluver17} is not shown in Figure \ref{fig:SFR} because KINGFISH galaxies were chosen to cover the full  range of galaxy types, luminosities and masses properties and local ISM environments rather than being magnitude-limited sample.}
\end{table*}

\subsection{Star Formation Efficiency}\label{SFE}

Star formation efficiency (SFE), defined as SFR/\Mh, quantifies the current rate of gas consumption, dividing the SFR by \HI\ mass and SFE is expected to depend on the stellar mass of a galaxy \citep{schiminovich10, huang12, wong16, lutz17}. SFE and its inverse, the depletion time, have thus been commonly used to quantify gas consumption and test models of the stability of the galactic disks \citep[e.g.][]{wong16}.
 
\begin{figure}[htb!]	
\includegraphics[width=3.3in,keepaspectratio]{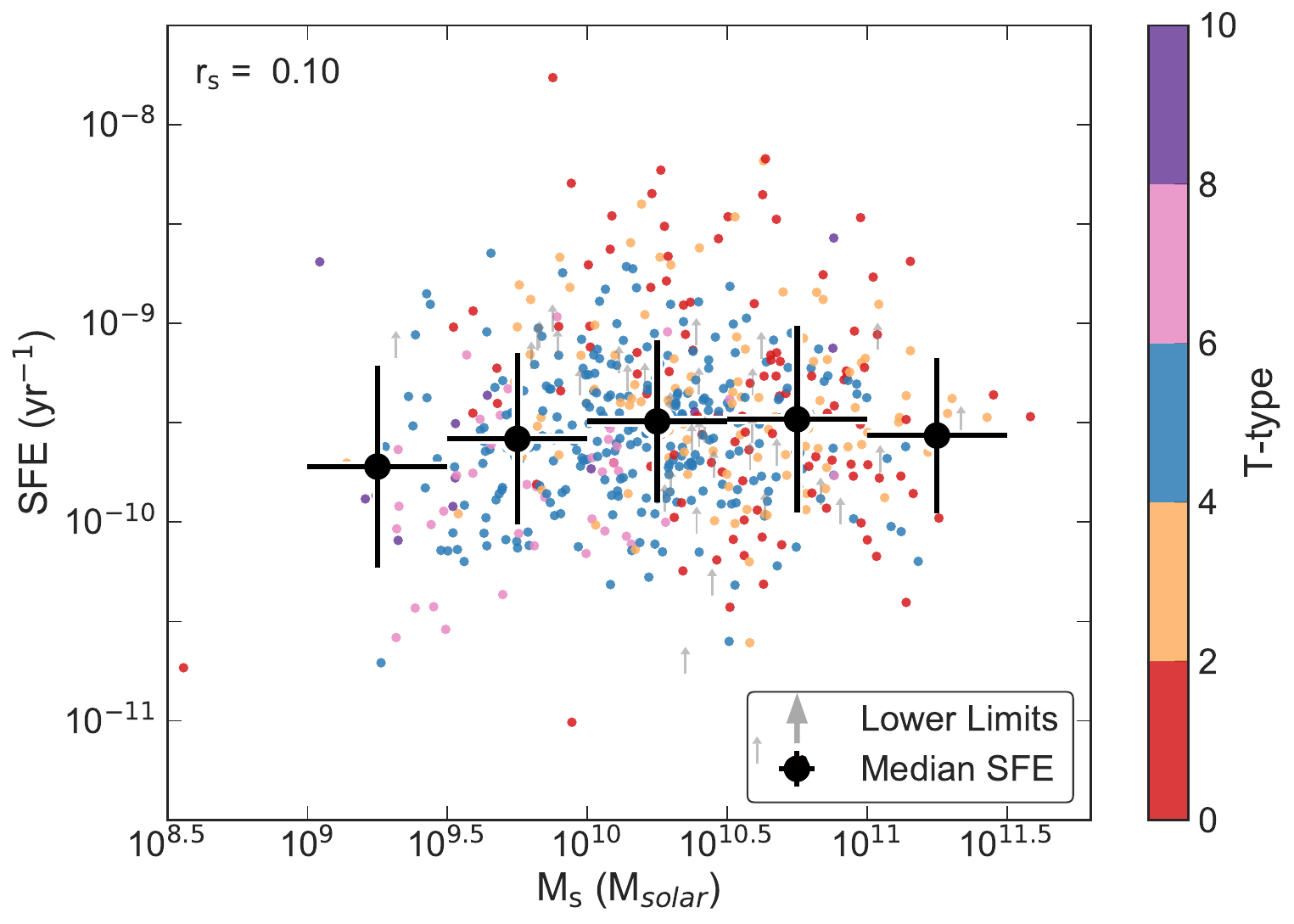}
	\caption{The SFE and the stellar mass for the \spiral. Median SFEs for each stellar mass bin are estimated including the \HI\ mass upper limits for non-\HI\ detections, and the SFE medians and $1\sigma$ values are listed in Table \ref{table:SFE_med}. For stellar masses ranging from 10$^{9}$ to 10$^{11.5}$ \Msolar, the SFE for spirals remains constant at a median value of $\log\; \mathrm{SFE} = -9.57$ and $1\sigma$ spread of $0.4$~dex. SFE appears to be almost constant with T-type and stellar mass, despite the fact high mass galaxies with T-type$<$2 fall below the MS and may have commenced quenching.}
	\label{fig:SFEMs_Ms}
\end{figure}

To investigate the relationship between star formation and \HI\ mass within the \spiral, we plot in Figure \ref{fig:SFEMs_Ms} SFE as a function of stellar mass. In Table \ref{table:SFE_med} we provide the median SFEs as a function of stellar mass, with the upper limits on the \HI\ mass being used for the \HI\ non-detections. The SFE remains relatively constant at a median SFE $=$ 10$^{-9.57}$ yr$^{-1}$, with a $1\sigma$ scatter of 0.44 dex for spiral galaxies with stellar masses between 10$^{9.0}$ and 10$^{11.5}$ \Msolar. While we see evidence for mass quenching in high stellar mass spirals in Figure \ref{fig:SFR}, the SFE appears to be constant for spiral galaxies falling on the MS and spiral galaxies that have (potentially) commenced quenching.

\begin{figure}[ht!]
	\center
	\includegraphics[width=3.3in,keepaspectratio]{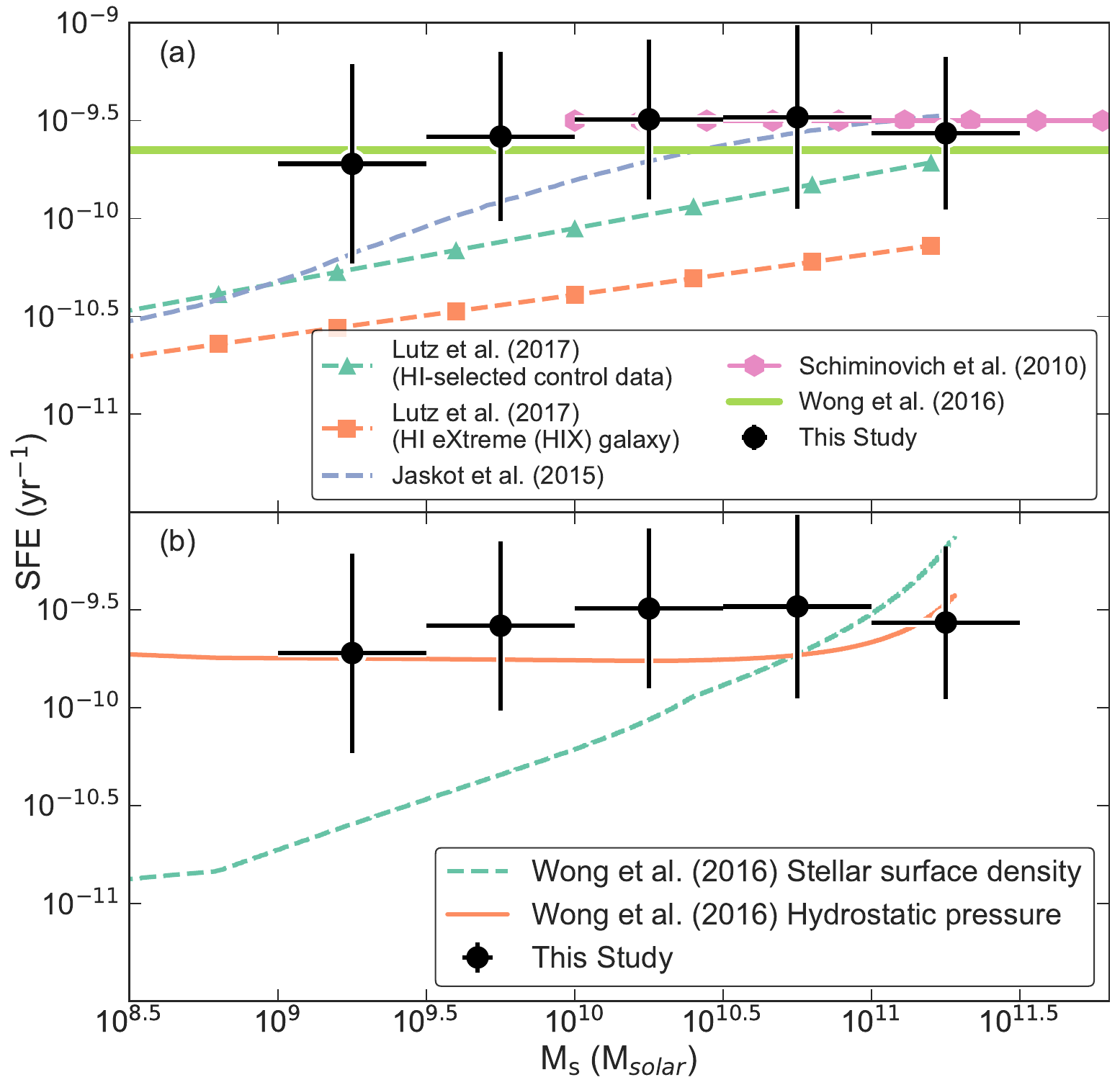}
	\caption{The median SFE and stellar mass for the spiral sample, alongside scaling relations observed in the prior literature \citep[i.e.,][]{schiminovich10, huang12, wong16, lutz17} (panel a) and two models presented by \citet{wong16} (panel b). \HI\ selected samples find that SFE increases with stellar mass, but this is an artifact of excluding low \HI\ mass galaxies. Stellar mass-selected samples, on the other hand, find that SFE is constant with stellar mass, with normalization between 10$^{-9.65}$~yr$^{-1}$ to 10$^{-9.5}$~yr$^{-1}$. Of the two models presented by \citet{wong16}, the hydrostatic pressure model provides the best agreement with our data.}
	\label{fig:SFEMs_Ms_comp}
\end{figure}

SFE versus stellar mass relations for both the \spiral\ and previous studies \citep{schiminovich10,jaskot15,wong16,lutz17} are compared in Figure \ref{fig:SFEMs_Ms_comp}a. \HI\ selected samples \citep{jaskot15, lutz17} exhibit an increasing SFE with stellar mass, however this reflects their selection bias against galaxies with low \HI\ masses. By contrast, stellar mass-selected samples \citep{schiminovich10,wong16} find that SFE is constant with stellar mass; however, previous studies have measured differing values of this constant, ranging from 10$^{-9.65}$ yr$^{-1}$ \citep{wong16} to 10$^{-9.5}$ yr$^{-1}$ \citep{schiminovich10}. \citet{wong16} provide two theoretically motivated relations for SFE versus stellar mass, which are both plotted in Figure \ref{fig:SFEMs_Ms_comp}b. The first relation assumes that the molecular gas fraction depends only on the stellar surface mass density, while the second assumes that this fraction depends on the hydrostatic pressure. Between stellar masses of 10$^{9.0}$ and 10$^{11.5}$ \Msolar, the hydrostatic pressure model of \citet{wong16} (which gives a constant SFE) shows the greatest consistency with our work and other stellar mass-selected samples.

\begin{table}[htb!]
\center
\caption{The median SFE and 1$\sigma$ values of each stellar mass bin for the \spiral\ shown in Figure \ref{fig:SFEMs_Ms}.}
\label{table:SFE_med}
\begin{tabular}{@{}cccc@{}}
\toprule
log(\Ms) & Median log(SFE) & 1$\sigma$& N      \\ 
(\Msolar) & (yr$^{-1}$)  & (dex) &  \\ \hline
9.25 & -9.72 &0.51& 29\\
9.75 & -9.58& 0.43& 128\\
10.25 & -9.49& 0.41 & 222\\
10.75 & -9.48 & 0.47&164\\ 
11.25 & -9.56 &0.39 &35\\ \hline
\end{tabular}
\end{table}

\section{Discussion}\label{Discussion}

\subsection{Is There an Upper Limit to the \Mh?}
For the \spiral, we find that \HI\ mass increases with stellar mass, from 10$^{9.14}$ \Msolar\ at a stellar mass of 10$^{9.25}$ \Msolar\ to 10$^{9.93}$ \Msolar\ at a stellar mass of 10$^{11.25}$ \Msolar\ (see Figure \ref{fig:MhiMs_spiral}). We also observe a stellar mass-dependent upper limit on \HI\ mass. In this section, we discuss the reason for this upper limit.

Both our \HI\ and stellar mass-selected samples imply an upper limit for \HI\ mass as a function of stellar mass, and such thresholds are also seen in prior literature \citep[e.g.,][]{maddox15}. Is this upper limit for \HI\ mass expected from theory? \citet{maddox15} argue that the maximum \HI\ fraction for galaxies with stellar masses $>$ 10$^9$ \Msolar\ is set by the upper limit in the halo spin parameter, $\lambda$. The halo spin parameter is defined as 
\begin{eqnarray}
\lambda \equiv J_{\mathrm{halo}}E_{\mathrm{halo}}^{1/2}G^{-1}M_{\mathrm{halo}}^{-5/2},
\end{eqnarray}
where $J_{\mathrm{halo}}$ is the galaxy halo's angular momentum, $E_{\mathrm{halo}}$ its total energy, and $M_{\mathrm{halo}}$ its total mass \citep{boissier00}. \citet{maddox15} determine the halo spin parameter of the ALFALFA galaxies and find that at a fixed stellar mass, the galaxies with the largest \HI\ mass also have the largest halo spin parameter (see their Figure 6). The large halo spin of a galaxy stabilizes the high \HI\ mass disk, preventing it from collapsing and forming stars.

\citet{maddox15} also measure the largest spin parameter to be $\lambda$ $\sim$ 0.2, confirming the upper limit on the halo spin parameter predicted by numerical $N$-body simulations of cold dark matter \citep{knebe08}. They conclude that the upper limit on \HI\ fraction is set by the upper limit of the halo spin parameter due to
the empirical correlation between the halo spin parameter and the \HI\ fraction.

\citet{obreschkow16} finds that for isolated local disk galaxies the fraction of atomic gas, $f_{\mathrm{atm}}$, is described by a stability model for flat exponential disks. To see if the observed upper limit to the \HI\ fraction of the \spiral\ can be explained by the upper limit of the halo spin parameter, we calculate the $f_{\mathrm{atm}}$ relationship for $\lambda$ $\approx $ 0.112, following the method outlined in \citet{obreschkow16}. Though $\lambda$ $\sim$ 0.2 is the maximum spin of a spherical halo, we choose to calculate the $f_{\mathrm{atm}}$ at $\lambda$ $\approx $ 0.112 because 99\% of galaxy halos are predicted to lie below $ \lesssim $ 0.112 \citep{bullock01}. We define the fraction of atomic gas as
\begin{eqnarray} \label{eq:simple}
f_{\mathrm{atm}} = \frac{1.35M_{H{\sc I}}}{M},
\end{eqnarray}
where M is the disk baryonic mass ($M$ $=$ \Ms $+$ 1.35$M_{H{\sc I}}$) and the factor of 1.35 accounts for the universal helium fraction \citep{obreschkow16}. \citet{obreschkow16} models the rotation curve of spiral galaxies as
\begin{eqnarray}\label{eq:fatm}
f_{\mathrm{atm}} = \min \left \{1, 2.5q^{1.12} \right \},
\end{eqnarray}
where $q$ is the global stability parameter. This parameter is defined as:
\begin{eqnarray}\label{eq:q}
q = \frac{j\sigma}{GM} = 0.22\frac{\lambda}{0.03}\left (\frac{M}{10^9M_\odot} \right )^{-1/3},
\end{eqnarray}
where $j$ is the baryonic specific angular momentum of the disk, $\sigma$ is the velocity dispersion of the atomic gas and mass is in units of 10$^9$ \Msolar.
The global stability parameter is simplified by making two assumptions: firstly, that disk galaxies condense out of scale-free cold dark matter halos, and secondly, that $j$ $\propto$ $\lambda$ \citep{obreschkow14, obreschkow16}.
Under these assumptions, Equation \eqref{eq:fatm} simplifies to:
\begin{equation}
f_{\mathrm{atm}} = \min \left \lbrace 1, 0.5 \left (\frac{\lambda}{0.03}\right )^{1.12}\left (\frac{M}{10^9M_\odot} \right )^{-0.37} \right \rbrace,
\end{equation}
and using Equation \eqref{eq:simple}, this can be rearranged to give:
\begin{align}\label{eq:mhi}
\frac{M_{H{\sc I}}}{10^9M_\odot} = \min \Biggl \lbrace &\frac{M}{1.35\times 10^9M_\odot}, \nonumber \\
&18\lambda^{1.12}\left (\frac{M}{10^9M_\odot} \right )^{-0.63}  \Biggr \rbrace.
\end{align}

\begin{figure}[htb!]
	\includegraphics[width=3.3in,keepaspectratio]{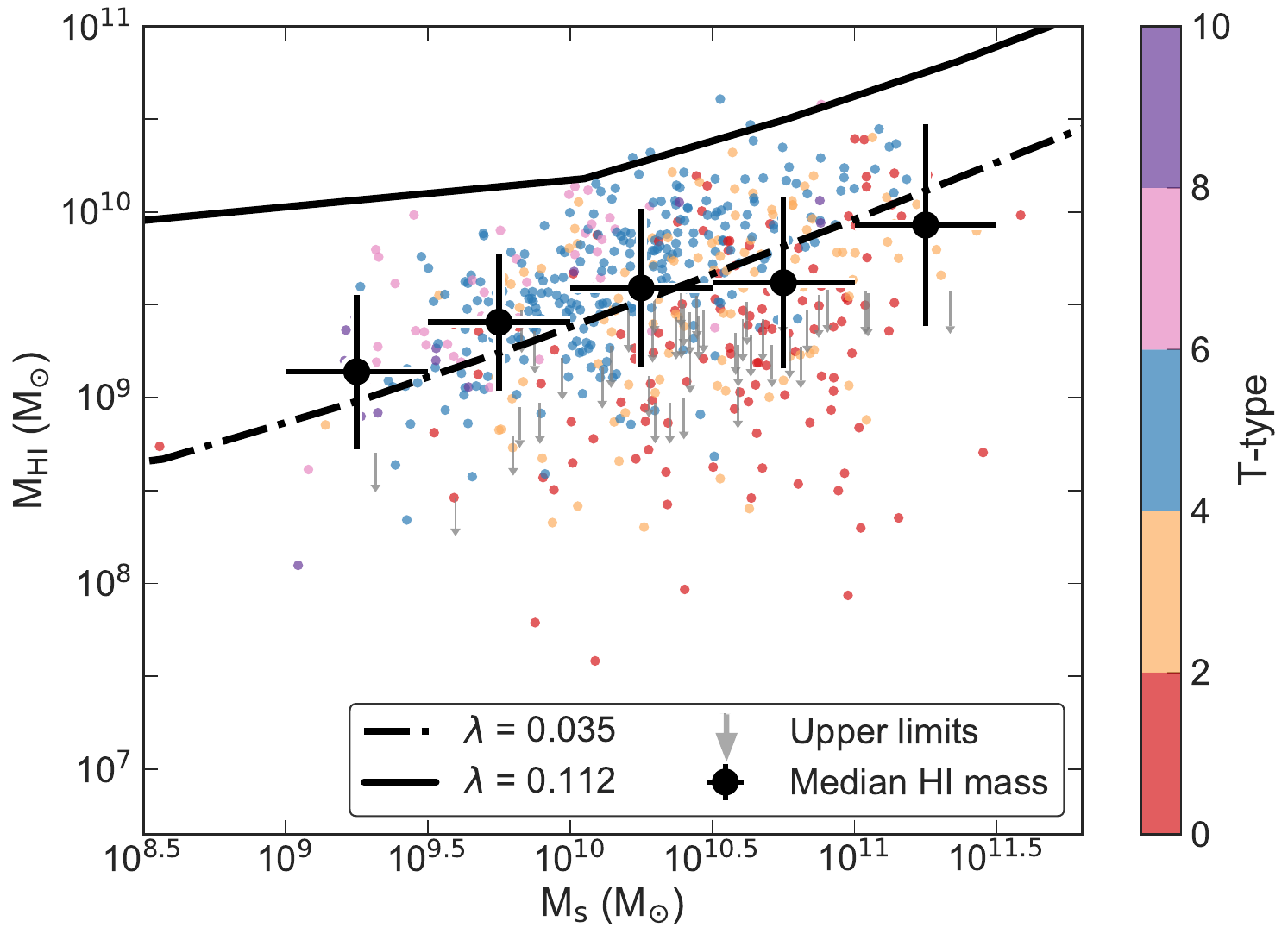}
	\caption{ \HI\ mass versus stellar mass for the \spiral. The solid and dashed black lines represent the model, Equation \eqref{eq:mhi}, for when $\lambda$ $=$ 0.112 and 0.035, respectively. The median bins of the \spiral\ are in good agreement with the model of \citet{obreschkow16}, as they lie on the expected mean $f_{\mathrm{atm}}$ (for $\lambda$ $=$ 0.035). The upper \HI\ masses of the \spiral\ line up with the expected $f_{\mathrm{atm}}$ for $\lambda$ $=$ 0.112, and therefore the maximum \HI\ mass for a given stellar mass is determined by the upper limit in the halo spin parameter.}
	\label{fig:halospin}
\end{figure}

Figure \ref{fig:halospin} illustrates the comparison between Equation \eqref{eq:mhi} and the empirical \HI\---stellar mass distribution of the \spiral. We also include the predicted $f_{\mathrm{atm}}$ curve for $\lambda$ $\approx $ 0.03, as this value of $\lambda$ corresponds to the mode of the empirically measured halo spin parameter distribution \citep{bullock01}. The median bins of the \spiral\ are in agreement with this prediction of $f_{\mathrm{atm}}$, while the highest \HI\ mass galaxies lie below the predicted upper limit for $f_{\mathrm{atm}}$, when $\lambda = 0.112$. We find that the model of \citet{obreschkow16} matches well with the empirical data of the \spiral, consistent with the hypothesis that the upper limit of the \HI\ fraction is set by that of the halo spin parameter.
 
\subsection{Why is SFE Constant?}
We find that star formation efficiency is constant across two orders of magnitude of stellar mass, which agrees with the findings of \citet{catinella10} and \citet{wong16}, while disagreeing with others \citep[e.g.,][]{huang12, lutz17}. \citet{wong16} tested two models for molecular gas content within galaxies: one where molecular gas is a function of stellar surface density and another where it is a function of hydrostatic pressure. The stellar surface density prescription \citep{leroy08,zheng13} defines the molecular-to-atomic ratio, $R_{\mathrm{mol}}$, as
 \begin{eqnarray}\label{rmols}	
R_{\mathrm{mol},s} = \frac{\Sigma_\ast}{81 M_\odot \, \mathrm{pc}^{-1}},
\end{eqnarray}
where $\Sigma_{\ast}$ is the stellar surface density. The $R_{\mathrm{mol}}$ for the hydrostatic pressure prescription \citep{zheng13} is defined as
 \begin{eqnarray}\label{rmolph}	
R_{\mathrm{mol},p} =\left( \frac{P_h}{1.7\times10^4 \, \mathrm{cm}^{-3} \, \mathrm{K} \, k_B}\right )^{0.8},
\end{eqnarray}
where $P_h$ is the hydrostatic pressure \citep{elmegreen89} and $k_B$ is the Boltzmann constant.

Similarly to \citet{wong16}, we find that the constant SFE can be described by a model of the marginally stable disk, while the hydrostatic pressure model provides a better prescription for estimating the SFE and molecular-to-atomic ratio. For massive galaxies with large optical disks, previous studies \citep[e.g.,][]{leroy08,wong13} observed a correlation between $R_{\mathrm{mol}}$ and stellar surface density. The two models given by Equations \eqref{rmols} and \eqref{rmolph} also predict similar $R_{\mathrm{mol}}$ and integrated SFE for high mass galaxies \citep{wong16}. But for low mass galaxies, the stellar surface density prescription does not predict the observed SFE because this prescription is unable to convert the \HI\ to molecular hydrogen, and underestimates the amount of molecular hydrogen in regions with low stellar surface densities. Therefore, this method underestimates the SFR and SFE in dwarf galaxies. While the stellar surface density model predicts that SFE will decrease for smaller stellar mass galaxies, the hydrostatic pressure model predicts a higher molecular hydrogen content for low mass galaxies, and therefore a constant SFE with stellar mass, agreeing with the empirical data.

We note that \citet{obreschkow16} predict that most of the baryons in dwarf galaxies are in the form of \HI\ gas ($f_{\mathrm{atm}}$ $=$ 1) and therefore have low SFE because these systems are inefficient at converting their \HI\ gas to molecular gas. We do not observe a decrease in SFE because these galaxies are below the stellar mass range probed in our \spiral. Our lowest stellar mass bin of 10$^{9.0}$ and 10$^{9.5}$ \Msolar\ hints at a turn-over for the low mass dwarf, as predicted by \citet{obreschkow16}. Combining the next generation of \HI\ survey, WALLABY, with 20-cm radio continuum from ASKAP, we will measure the \HI\ properties and SFR of $\sim$ 600,000 galaxies, including a large number of dwarf galaxies populating the low mass end of Figure \ref{fig:SFEMs_Ms_comp}. Observing these low stellar mass dwarfs will provide a more complete picture about \HI\ content and how efficiently dwarfs convert \HI\ to molecular gas. 

\section{Summary}\label{summary}
We have measured the relationship between \HI\ mass, stellar mass and SFE using HICAT, archival \HI\ data and new WISE photometry. For this work, we provide new WISE aperture photometry for 3,831 out of 4,315 sources of HICAT and created three samples, an \HI\ selected sample, a \spiral\ and a \Mss. We find the following:
\begin{enumerate}
  \item Sample selection and biases are critical when interpreting and comparing measured relationships between \HI\ mass, stellar mass and SFE. \HI\ selected samples often exclude \HI\ poor galaxies (unlike stellar mass selected samples) resulting in measurements of high median \HI\ masses and low median SFEs.
  \item \HI\ mass increases with stellar mass for the \spiral\ with a power-law index of 0.34. Also, at a given stellar mass, \HI\ mass increases with T-type and dispersion in \HI\ masses narrows for individual T-types. For example, for the $10^{10}$ \Msolar\ to $10^{10.5}$ \Msolar\ stellar mass bin, the median \HI\ mass and scatter is $10^{9.26}$ \Msolar\ and 0.59 dex for T-types 0 to 2 and the $10^{9.72}$ \Msolar\ and 0.31 dex for T-types 6 to 8.
  \item \HI\ mass is constant with stellar mass for the \Mss. While \HI\ mass increases with stellar mass for spiral galaxies, the fraction of elliptical galaxies with little \HI\ gas also increases with stellar mass, producing the observed flat relation. 
    \item The observed upper limit to the \HI-stellar mass distribution of the \spiral\ is consistent with the predicted \HI-stellar mass curve for the upper limit for the halo spin parameter ($\lambda$ $=$ 0.112). This is consistent with the hypothesis the maximum \HI\ fraction is set by that of the halo spin parameter.
      \item For a subsample of the \spiral\ with stellar mass $\le$ $10^{10.5}$ \Msolar, we measure the MS to be $\log SFR = 0.7 (\log M_{\ast} - 10) - 0.09$ with a $1\sigma$ scatter of 0.27 dex. We see evidence of mass quenching \citep[e.g.][]{kauff03} as the median SFR is constant for spiral galaxies with stellar masses $>$ $10^{10}$ \Msolar.
  \item For the \spiral, SFE is constant ($=$ $10^{-9.57}$ $\pm$ 0.44 yr$^{-1}$) for 2.5 orders of magnitude in stellar mass and agrees with comparable measurements of stellar mass selected samples of galaxies \citep{catinella10, schiminovich10}. This results is in broad agreement with the hydrostatic pressure model \citep{wong16}.  
	\item SFE is constant as a function of T-type and is constant for spiral galaxies that show evidence of mass quenching. 
\end{enumerate}

\acknowledgments
We would like to thank M. Cluver, T. Dolley, A. Groszek, V. Kilborn, K. Lutz, N. Maddox, G. Meurer, D. Obreschkow, and I. Wong for the useful and insightful discussions.

This publication makes use of data products from the Wide-field Infrared Survey Explorer, which is a joint project of the University of California, Los Angeles, and the Jet Propulsion Laboratory/California Institute of Technology, funded by the National Aeronautics and Space Administration. This research has made use of the NASA/IPAC Extragalactic Database (NED) which is operated by the Jet Propulsion Laboratory, California Institute of Technology, under contract with the National Aeronautics and Space Administration and the HyperLeda database (http://leda.univ-lyon1.fr). 


\bibliographystyle{apj}
\bibliography{Thesis}


\appendix
\startlongtable
\begin{deluxetable}{p{25mm}p{18mm}p{120mm}}
\tablecaption{HI-WISE parameter descriptions \label{tab:param_desc}}
\tablehead{
\colhead{Parameter }              & \colhead{Units}         & \colhead{Description}}
\startdata	        	      	   
Name                     &           &   HIPASS designation\\
RA                        & deg           & Right ascension (J2000)\\
Dec                      & deg           & Declination (J2000)\\
T-type & & The T-type of galaxies reported by \citet{bonne15} or NED.\\
Dist & Mpc & Luminosity distance\\
Dist Source Flag & & Source of the distance : 0. Cosmicflows-3 \citep{tully16}, 1. HICAT, 2. NED \\
Optical Match Type & &Our category matching choice: 0. no optical counterpart, 1. position and velocity match, 2. position match. \\
vel$_{HI}$        & km s$^{-1}$         & The flux weighted velocity average between minimum and maximum profile velocity.  \\
$     S_{\rm p}$       & Jy            & peak flux density of profile\\
$   S_{\rm int}$     & Jy  km s$^{-1}$   & integrated flux of source (within region $v_{\rm lo}$, $v_{\rm hi}$ and box size)\\
W1  & mag& W1 isophotal magnitude    \\
W1err  & mag& W1 magnitude error \\
W2  & mag& W2 isophotal magnitude    \\
W2err  & mag& W2 magnitude error \\
W3 &mag & W3 `best' magnitude \\
W3err  & mag& W3 magnitude error \\
W3f && W3 `best' photometry type: 0. isophotal, 10. total magnitude \\
W4 &mag & W4 `best' magnitude \\
W4err  & mag& W4 magnitude error \\
W4f && W4 `best' photometry type: 0. isophotal, 10. total magnitude \\
R1iso   &    arcsec  &   W1 1-sigma isophotal radius (semi-major axis)\\
R2iso    &  arcsec    & W2 1-sigma isophotal radius or photometry aperture (semi-major axis)\\
R3iso     & arcsec	 &   W3 1-sigma isophotal radius or photometry aperture (semi-major axis)   \\
R4iso   &   arcsec   &  W4 1-sigma isophotal radius or photometry aperture (semi-major axis)\\
ba       &  &         axis ratio based on the W1 3-sigma isophote\\
pa       &  deg&        position angle (east of north) based on the W1 3-sigma isophote\\
W1W2    &   mag  &    	W1-W2 color, where the W1 aperture is matched to the W2 1-$\sigma$ isophotal aperture \\
W1W2err   &  mag &       W1-W2 color uncertainty\\
W2W3 &      mag   &     W2-W3 color, using the W2 isophotal aperture and the W3 isophotal aperture\\
W2W3err    & mag    &    W2-W3 color uncertainty\\
W1-W2    &  mag   &     K-corrected (rest-frame) W1-W2 color\\
W2-W3   &   mag      &   K-corrected (rest-frame)   W2-W3 color\\
log $L_{W1}$ &log L$_{\odot}$ & W1 $\nu L_\nu$ luminosity (log10) \\
log $L_{W1}$ err&log L$_{\odot}$ & uncertainty in log $L_{W1}$ \\
log $L_{W2}$ &log L$_{\odot}$ & W2 $\nu L_\nu$ luminosity (log10) \\
log $L_{W2}$ err&log L$_{\odot}$ & uncertainty in log $L_{W2}$ \\
log $L_{W3}$ &log L$_{\odot}$ & W3 $\nu L_\nu$ luminosity (log10). This includes the stellar continuum.  \\
log $L_{W3}$ err&log L$_{\odot}$ & uncertainty in log $L_{W3}$ \\
log $L_{W4}$ &log L$_{\odot}$ & W4 $\nu L_\nu$ luminosity (log10).This includes the stellar continuum.   \\
log $L_{W4}$ err&log L$_{\odot}$ & uncertainty in log $L_{W4}$ \\
log $L_{W1}(L_\odot)$ &log L$_{\odot}$ & W1 in-band luminosity (log10) \\
log $L_{W1}(L_\odot)$ err&log L$_{\odot}$ & uncertainty in log $L_{W1}(L_\odot)$ \\
W1flux &   mJy  &       WISE W1 K-corrected flux\\
W2flux   &   mJy  &       WISE W2 K-corrected flux\\
w3PaH  &   mJy   &     WISE W3 K-corrected  flux, with the stellar continuum subtracted;  result is the ``PaH" flux\\
 w4dust &     mJy     &   WISE W4 K-corrected  flux, with the stellar continuum subtracted;  result is the warm dust flux    \\
log \Ms  & log(\Msolar)& Stellar mass (log10), based on the W1-W2 K-corrected color and the W1 in-band luminosity ($L_{W1}(L_\odot)$) \citep{cluver14}.   \\
log \Ms\ err  & log(\Msolar)& Uncertainty in  log \Ms  \\
log \Mh  &  log(\Msolar)& \HI\ mass (log10)  \\
log \Mh\ err  & log(\Msolar)& Uncertainty in log \Mh  \\
log SFR & log(\Msolar yr$^{-1}$)&  Star formation rate (log10) based on the W3 $\nu L_\nu$ luminosity \citep{brown17}. \\
log SFR err& log(\Msolar yr$^{-1}$)&  Uncertainty in log SFR   \\
log SFE & log(yr$^{-1}$)&  Star formation efficiency (log10). \\
log SFE err& log(yr$^{-1}$)&  Uncertainty in log SFE   \\
\enddata
\tablecomments{Table 8 is published in its entirety in the electronic 
edition of the {\it Astrophysical Journal}.  A portion is shown here 
for guidance regarding its form and content.}
\end{deluxetable}

Table \ref{tab:param_desc} provides a full list of parameters of HICAT+WISE (HI-WISE) catalog which is available in machine-readable format.

\end{document}